\documentclass[final,3p,times]{elsarticle}

\usepackage{amssymb}

\usepackage[utf8]{inputenc}
\usepackage{subfig}
\usepackage{enumitem}
\usepackage{siunitx}
\usepackage{amsmath,xcolor}
\usepackage{nicefrac}
\usepackage{hyperref}
\hypersetup{
    colorlinks=true,
    linkcolor=blue,
    filecolor=magenta,      
    urlcolor=cyan,
}
\graphicspath{{./bilder/}}

\newcommand{\fenics}{FEniCS}

\newcommand{\coma}{\>\text{,}}
\newcommand{\point}{\>\text{.}}
\newcommand{\nv}{\text}
\newcommand{\ve}[1]{\boldsymbol{#1}} 
\newcommand{\te}[1]{\boldsymbol #1} 
\newcommand{\T}{^\top}
\newcommand{\dv}{\,\text{d}V}
\newcommand{\da}{\,\text{d}A}
\newcommand{\dx}{\,\text{d}x}
\newcommand{\elli}{{\ell_\text{i}}}
\newcommand{\ellc}{{\ell_\text{c}}}
\newcommand{\gc}{\mathcal{G}_\text{c}}
\newcommand{\gcb}{\mathcal{G}_\text{c}^\text{b}}
\newcommand{\gci}{\mathcal{G}_\text{c}^\text{i}}
\newcommand{\gb}{\mathcal{G}^\text{b}}
\newcommand{\gi}{\mathcal{G}^\text{i}}
\newcommand{\gcia}{\mathcal{G}_\text{c}^\text{i,act}}
\newcommand{\gcih}{\hat{\mathcal{G}}_\text{c}^\text{i}}
\newcommand{\jt}{\mathcal{J}^\nv{tip}}

\newcommand{\gammai}{\Gamma^\text{i}}
\newcommand{\gammac}{\Gamma^\text{c}}
\newcommand{\gammalc}{\Gamma^{\ell_\text{c}}}
\newcommand{\gammali}{\Gamma^{\ell_\text{i}}}
\newcommand{\rheavi}{\mathcal{G}_\text{c}^\text{H}}
\newcommand{\rgauss}{\mathcal{G}_\text{c}^\text{G}}
\newcommand{\rexpo}{\mathcal{G}_\text{c}^\text{E}}
\newcommand{\reheavi}{E^\text{H}}
\newcommand{\rtan}{E^\text{T}}
\newcommand{\phii}{\varphi_\text{i}}

\DeclareMathOperator{\sign}{sign}
\DeclareMathOperator{\tr}{tr}

\renewcommand{\Omega}{\varOmega}
\renewcommand{\Gamma}{\varGamma}
\renewcommand{\Sigma}{\varSigma}
\renewcommand{\Psi}{\varPsi}

\newcommand{\legline}[1]{\raisebox{-0.1cm}{\protect\includegraphics{line#1.pdf}}}
\newcommand{\legpoint}[1]{\raisebox{-0.1cm}{\protect\includegraphics{point#1.pdf}}}

\usepackage{multirow}
\usepackage{booktabs}
\usepackage{mdframed}
\usepackage{mathtools}

\usepackage{listings}

\definecolor{gnupl-darkgreen}{HTML}{006400}
\definecolor{gnupl-darkred}{HTML}{8b0000}
\newcommand{\cex}[1]{Code snippet~\ref{#1}}

\journal{Eng. Fract. Mech.\qquad \texttt{{\normalfont \textcolor{red}{https://doi.org/10.1016/j.engfracmech.2020.107004 (added on May 05, 2020)}}}}

\begin{document}

\begin{frontmatter}



\title{Phase-Field Modeling of Crack Branching and Deflection in Heterogeneous Media}

\author[label1]{Arne Claus Hansen-Dörr}
\author[label1]{Franz Dammaß}
\author[label3]{René de Borst}
\author[label1,label2]{Markus Kästner\corref{cor1}}

\address[label1]{Institute of Solid Mechanics, TU Dresden, 01062 Dresden, Germany}
\address[label3]{University of Sheffield, Department of Civil and Structural Engineering, Mappin Street, Sir Frederick Mappin Building, Sheffield S1 3JD, UK}
\address[label2]{Dresden Center for Computational Materials Science (DCMS), TU Dresden, 01062 Dresden, Germany }
\cortext[cor1]{markus.kaestner@tu-dresden.de}

\begin{abstract}
This contribution presents a diffuse framework for modeling cracks in heterogeneous media. Interfaces are depicted by static phase-fields. This concept allows the use of non-conforming meshes. Another phase-field is used to describe the crack evolution in a regularized manner.

The interface modeling implements two combined approaches. Firstly, a method from the literature is extended where the interface is incorporated by a local reduction of the fracture toughness. Secondly, variations of the elastic properties across the interface are enabled by approximating the abrupt change between two adjacent subdomains using a hyperbolic tangent function, which alters the elastic material parameters accordingly.

The approach is validated qualitatively by means of crack patterns and quantitatively with respect to critical energy release rates with fundamental analytical results from Linear Elastic Fracture Mechanics, where a crack impinges an arbitrarily oriented interface and either branches, gets deflected or experiences no interfacial influence. The model is particularly relevant for phase-field analyses in heterogeneous, possibly complex-shaped solids, where cohesive failure in the constituent materials as well as adhesive failure at interfaces and its quantification play a role.
\end{abstract}


\end{frontmatter}

\section{Introduction}
\label{sec:intro}

\begin{figure}
\begin{mdframed}
    \begin{center}
    \textbf{Nomenclature}\\
    \ \\
    \begin{minipage}[b]{0.48\textwidth}
    \begin{tabular}{ll}
        \multicolumn{2}{l}{Latin symbols} \\
        $a$ & domain measure \\
        $A$ & cross-sectional area  \\
        $b$ & domain measure \\
        \multirow{2}{*}{$c$} & \multirow{2}{*}{$\left\lbrace \begin{matrix*}[l] \text{\,domain measure} \\ \text{\,phase-field}  \end{matrix*} \right.$}\\ 
           & \\ 
        $\mathcal{C}$ & circular integration domain \\
        $d$ & signed distance \\
        $D$ & physical dimensions \\
        $D_i$ & denominators for projection tensors \\
        $e_i$ & eigenvalues of the strain tensor $\te{\varepsilon}$\\
        $E$ & \textsc{Young}'s modulus \\
        $g$ & degradation function \\
        $\tilde{g}_i$ & configurational force \\
        $\mathcal{G}$ & energy release rate\\
        $\mathcal{G}_\text{c}$ & fracture toughness\\
        $I$ & functional \\
        $I_1, I_2, I_3$ & principal invariants of $\te{\varepsilon}$  \\
        $\mathcal{J}_i$ & crack driving force \\
        $\ell$ & regularization length scale\\
        $\boldsymbol{M}_i$ & projection tensors \\
        $n_i$ & surface normal vector\\
        $O$ & \multirow{3}{*}{$\left.\begin{matrix} \mbox{}\\\mbox{} \\\mbox{} \end{matrix} \right\rbrace$ integration constants} \\
        $P$ & \\
        $Q$ & \\
        $r$ & radius \\
        $s,t$ &  coordinates oriented to interface \\
        $t$ & time \\
        $\bar{t}_i$ & given surface traction \\
        $u_i,\bar{u}_i$ & displacement, given displacement \\
        $\ve{u}$ & displacement boundary condition \\
        $V$ & volume \\
        $x_i$ & index notation of coordinates \\
        $\bar{\ve{x}}=[\bar{x}\,\bar{y}]\T$ &  position of virtual crack tip \\
        $x,y$ &  cartesian coordinates\\
        & \\
        \multicolumn{2}{l}{Greek symbols} \\
        $\alpha$ & First \textsc{Dundurs}' parameter\\
        $\gamma$ & surface density\\
        $\varGamma$ & surface\\
        $\delta$ & \textsc{Dirac} distribution\\
        $\delta\bullet$ & test function for $\bullet$\\
        $\delta_{ij}$ & \textsc{Kronecker} delta \\
        $\Delta$ & increment of \dots\\
        $\te{\varepsilon}$, $\varepsilon_{ij}$ & strain tensor\\
        $\eta$ & residual stiffness \\
        $\eta_\text{f}$ & kinetic fracture parameter/viscosity \\
        $\kappa$ & external volume micro force\\
        $\nu$ & \textsc{Poisson} ratio
    \end{tabular}
    \end{minipage}%
    \begin{minipage}[b]{0.48\textwidth}
    \begin{tabular}{ll}
        $\xi_i$ & micro force traction\\
        $\pi$ & internal volume micro force\\
        $\sigma_{ij}$ & stress tensor\\  
        $\varSigma_{ij}$ & energy momentum stress tensor \\
        $\varphi$ & inclination angle\\
        $\psi$ & free energy density\\
        $\Psi$ & free energy\\
        $\varOmega$  & domain \\
    & \\
    \multicolumn{2}{l}{Sub-/superscripts} \\
        $+$ & tensile part\\
        $-$ & compressive part\\
        \hspace{1pt}\large\textasciicircum & compensated \dots \\
        $0$ & initial \dots\\
        $1,2$ & material numbering \\
        act & actual \dots \\
        b & bulk \dots \\
        BC & boundary condition\\
        c & crack \dots \\
        def & deformation \dots\\
        dis & dissipation \dots\\
        el & elastic \dots \\
        E & exponential description \\
        G & \textsc{Gaussian}-like description \\
        \multirow{2}{*}{H} & \multirow{2}{*}{$\left\lbrace \begin{matrix*}[l] \text{\,\textsc{Heaviside}-like description} \\ \text{\,sharp \textsc{Heaviside} jump}  \end{matrix*} \right.$}\\ 
           & \\ 
        i & interface \dots\\
        $\ellc$ & regularized crack/phase-field \dots \\
        $\elli$ & regularized interface \dots \\
        len & length along interface \\
        max & maximum \dots \\
        min & minimum \dots \\
        modE & varying \textsc{Young}'s modulus \dots \\
        $n$ & increment number \\
        ref & reference value \\
        s & side \\
        $t$ & natural boundary condition \\
        th & threshold \\
        tip & crack tip \dots \\
        $\top$ & transposed \\
        T & hyperbolic tangent regularization\\
        $u$ & essential boundary condition \\
        & \\
        \multicolumn{2}{l}{Abbreviations} \\
        FEniCS & open-source finite element package \\
        LEFM & Linear Elastic Fracture Mechanics \\
        PETSc & numerics library \\
        UFL & Unified Form Language\\
        &
    \end{tabular}
    \end{minipage}
    \end{center}
\end{mdframed}
\end{figure}

Crack propagation is one of the most severe mechanisms compromising the bearing capacity of engineering structures. The phase-field approach to fracture has proven to be a powerful tool for the numerical prediction of crack propagation. The method allows for the description of complex failure mechanisms, such as crack nucleation and arrest, as well as branching and merging phenomena \cite{bourdin_variational_2008,miehe_thermodynamically_2010, miehe_phase_2010, kuhn_continuum_2010}. 
The concept is based on the variational approach to brittle fracture \cite{francfort_revisiting_1998}, which is consistent with the energetic criterion of Griffith \cite{griffith_phenomena_1921}.
The key idea of the phase-field method is the regularization of the underlying energy functional~\cite{bourdin_numerical_2000}: Cracks are approximated by an auxiliary field, often referred to as the crack phase-field.
The phase-field variable continuously varies from the intact to the fully broken material state; cracks are regularized using a finite length scale $\ellc$. Furthermore, the approach allows for the description of cracks with a non-conforming mesh, i.e. the element edges do not have to be aligned with the crack.

Modern engineering materials often consist of several components, e.g. fiber-reinforced composites.
As a separation of these components can occur, the adhesive interfaces within a heterogeneous material can significantly influence the mechanical behavior of structures under external loading.
Therefore, it is indispensable to account for interfaces in numerical simulations of fracture phenomena.

In the context of the Linear Elastic Fracture Mechanics (LEFM) analyses of He and Hutchinson~\cite{he_crack_1989,he_kinking_1989}, the interface is defined as a zone of infinitesimal width, which is assigned a fracture toughness that differs from the bulk material. Different setups were investigated, where a crack impinges a possibly inclined interface and either experiences no interfacial influence regarding the crack path or gets deflected. These fundamental and insightful investigations serve as analytical reference for numerical models, which incorporate interfaces in different manners. 

An approach quite close to the LEFM view on the interface was proposed by Paggi and Reinoso~\cite{paggi_revisiting_2017} and later Guillén-Hernández et al.~\cite{guillen-hernandez_micromechanical_2019}. They introduced a phase-field model for brittle fracture, where the interface is captured by cohesive zone elements: The cohesive zone approach is extended so that the phase-field affects both, the bulk material and interface stiffness. Good agreement of their hybrid model and the analytic investigations of He and Hutchinson~\cite{he_crack_1989} was achieved. A drawback of the model is the necessity of a mesh-conforming interface, i.e. cohesive zone elements have to be introduced along the interface. An extension to moving interface problems as apparent in phase transitions is thus not straightforward. 

Kuhn and Müller~\cite{kuhn_phase_2019} recently presented an approach where the interface is incorporated by adding a phase-field dependent surface term to the total energy of the elastic body. A qualitative study demonstrates the general capability of the modeling approach. For now, their formulation relies on a conforming interface description similar to the cohesive zone approach above.

Nguyen et al.~\cite{nguyen_phase-field_2016,nguyen_phase_2019} proposed a phase-field model for interface failure, where the relations for a standard cohesive zone model are applied to a regularized interface. The regularization is inspired by the crack surface density and takes the same form. Instead of the fracture toughness, a cohesive energy depending on the regularized displacement jump captures the energetic contribution to the total energy. In contrast to a classical cohesive zone model, the interface can be described in a non-conforming manner using a level-set, which can be generated from CT images.

Schneider et al.~\cite{schneider_phase-field_2016} presented a multiphase-field model capable of depicting cracks along the interface separating solid phases. By using a multiphase-field model, the interface is accounted for by a grain boundary energy, which is, however, different from the fracture toughness. For a varying grain boundary energy, the authors reproduced phenomena similar to those considered by He and Hutchinson~\cite{he_crack_1989}.
A quantitative comparison is not straightforward because the fracture toughness, present in~\cite{he_crack_1989}, was not used in~\cite{schneider_phase-field_2016} to characterize the fracture properties of the interface. The advantage of the multiphase-field model is the capability to describe non-conforming interfaces in a framework, which already allows for phase transitions and thus, a possible evolution of the interface itself. 

Hansen-Dörr et al.~\cite{hansen-dorr_numerical_2017,hansen-dorr_phase-field_2018,hansen-dorr_phase-field_2019} have presented the concept of a regularized, diffuse interface: In the context of the fracture phase-field, the regularized interface $\gammali$ is defined as a \textit{narrow} subdomain of a solid with a small --~but finite~-- characteristic width $\elli$, which is assigned an interface fracture toughness $\gci$.
As the length scales of the interface and crack interact, the effective fracture toughness of the interface depends on the characteristic length scales $\elli$ and $\ellc$, and on the fracture toughness of the surrounding bulk material.
In \cite{hansen-dorr_phase-field_2019}, a compensation of this effect by means of definition of a modified numerical interface fracture toughness was proposed. The advantage of this approach is simplicity, while keeping accuracy.
Once the compensation has been determined, a non-conforming interface can be embedded. This approach can be extended to moving interfaces, where evolution equations for the interface have to be implemented, cf.~\cite{mosler_novel_2014,kiefer_numerical_2017}.

The characteristic interface width $\elli$ is closely related to experimental work of Park and Chen~\cite{park_experimental_2011}, and Parab and Chen~\cite{parab_crack_2014}. In both papers, projectiles are fired at brittle solids to provoke dynamic crack propagation towards a perpendicular interface. The interface has a varying, finite width and is made of an adhesive, gluing two brittle solids together. Depending on the interface width, different fracture phenomena occur. The same behavior is observed in the present paper and underlines the fact, that the characteristic width of the interface is not a purely numerical parameter. 

This contribution extends the modeling approach developed by Hansen-Dörr et al.~\cite{hansen-dorr_phase-field_2019} to obtain a more numerically robust description of the interface and to incorporate elastic heterogeneities.
The first issue is addressed by introducing two continuous regularization functions that characterize the interface. These functions which smoothly describe the transition from the bulk to the interface fracture toughness, are considered instead of an actual material stripe assigned the interface fracture toughness.
Elastic heterogeneities are captured by postulating an approximation for the elastic constants in the interface region depending on the surrounding bulk materials properties.
The consequences of different interface regularizations are discussed in detail. The model is validated by qualitative and quantitative comparisons to analytical results from LEFM~\cite{he_crack_1989}.
In order to obtain a controlled crack growth through or along the interface, \textit{surfing boundary conditions}~\cite{hossain_effective_2014} are applied.
For a quantitative insight into the failure mechanisms, the concept of configurational forces is exploited~\cite{kuhn_discussion_2016,kuhn_energetic_2011}.

This paper is structured as follows. Section~\ref{sec:theory} introduces the phase-field model and the diffuse modeling of material heterogeneities like interfaces or dissimilar elastic materials. Furthermore, the compensation approach is outlined. A detailed quantitative and qualitative evaluation of the capabilities of the model and a comparison to four fundamental simulation setups from LEFM~\cite{he_crack_1989} is presented in Section~\ref{sec:numericresults}.
The paper concludes with a brief summary of the model and the results and a discussion.

\section{Phase-field modeling of regularized material heterogeneities}
\label{sec:theory}
\subsection{Introduction of crack surface density}
\begin{figure}[t]
	\centering
	\subfloat[Discrete crack representation]{\label{fig:disccrack}\includegraphics{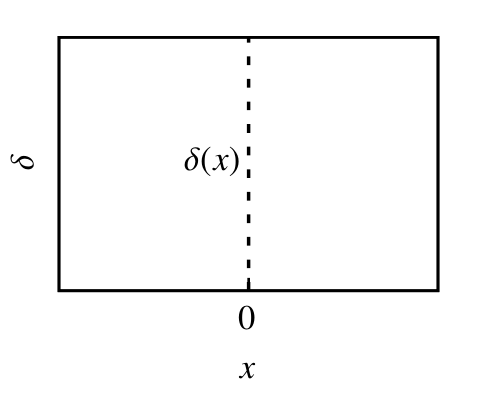}}
	\subfloat[Regularized crack representation]{\label{fig:regcrack}\includegraphics{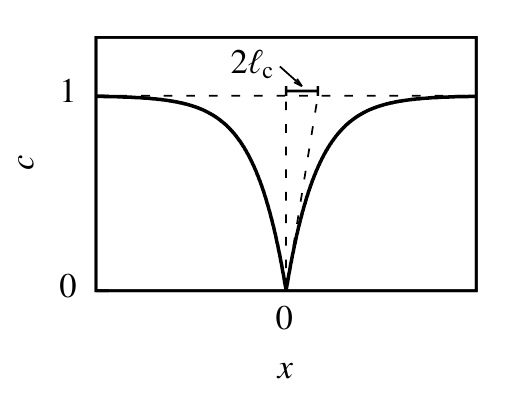}}
	\caption{In \textbf{(a)}, the location of the sharp crack is described by the \textsc{Dirac} distribution. The regularized representation using an exponential function is depicted in \textbf{(b)}. The length scale parameter $\ellc$ controls the width of the transition region from $c=0$ to $c=1$. 
	}
	\label{fig:discreg}
\end{figure}

The idea of phase-field modeling of fracture is the introduction of an additional scalar field $c\in\left[0,1\right]$, which implements a smooth transition from intact ($c=1$) to fully broken ($c=0$) material. The additional field $c$ is referred to as the phase-field in view of the resemblance of the concept to classical phase-field models. Suppose a one-dimensional rod with $x\in (-\infty,\infty)$ of cross-sectional area $A$ which is cracked at the center at $x=\SI{0}{\milli\meter}$: The crack location can be fixed using a \textsc{Dirac} distribution, cf. Figure~\ref{fig:disccrack}, and the total crack surface $\gammac$ can be obtained by integration over the domain
\begin{equation}
\label{eq:diracgamma}
    \gammac=\int\limits_{\Omega}\delta(x)\dv=\int\limits_{-\infty}^\infty \delta(x)A\dx=A\coma
\end{equation}
yielding the intuitive result $\gammac=A$. The motivation to describe the crack surface in a regularized manner arises in the context of finite element analyses. The smooth function $c$ enables the use of non-conforming meshes, which obviates the need for remeshing in case of crack propagation. Following Bourdin et al.~\cite{bourdin_variational_2008}, 
the \textsc{Dirac} distribution is regularized using an exponentially shaped function
\begin{equation}
\label{eq:1dregcrack}
    c(x)=1-\exp\left( \frac{-\vert x\vert}{2\ellc}\right)
\end{equation}
yielding a representation, which is depicted in Figure~\ref{fig:regcrack}. The characteristic length scale $\ellc$ controls the maximum gradient of the regularization, which has to be resolved in a finite element implementation. 

It has been shown~\cite{miehe_thermodynamically_2010} that a functional
\begin{equation}
    \label{eq:defCSD1d}
    I\left[c,c^\prime\right]=\int\limits_{\Omega} \frac{1}{4\ellc}\left[ \left(1-c\right)^2 +4\ellc^2\left( c^\prime\right)^2\right]\dv
\end{equation}
can be found, where Equation~\eqref{eq:1dregcrack} is the solution to the \textsc{Euler-Lagrange} equation for $I\left[c,c^\prime\right]\rightarrow\min$ and \mbox{$c^\prime=\text{d} c/\text{d} x$}, subject to the boundary condition $c^\prime(x\rightarrow\pm\infty)=0$. Note, that inserting Equation~\eqref{eq:1dregcrack} into \eqref{eq:defCSD1d}, in analogy to Equation~\eqref{eq:diracgamma}, yields $I=A=\gammalc$, which is why $\gammalc$ is used below instead of $I$. A three-dimensional generalization
\begin{equation}
    \label{eq:defCSD3d}
    \gammalc\left[c,c_{,i}\right]=\int\limits_{\Omega} \underbrace{
    \frac{1}{4\ellc}\left[ \left(1-c\right)^2 +4\ellc^2 c_{,i}c_{,i}\right]}_{\gamma^{\ellc}}\dv
\end{equation}
is obtained by replacing $c^\prime$ by the gradient, where $\gamma^\ellc$ is referred to as crack surface density, cf.~\cite{miehe_thermodynamically_2010}.
Herein, the summation convention and $(\bullet)_{,i}=\partial(\bullet)/\partial x_i$ apply.

\subsection{Governing differential equations and \textsc{Clausius-Duhem} inequality}
\label{sec:diffeqcdu}
The local form of the momentum balance, neglecting volume forces and inertia, reads
\begin{equation}
\label{eq:linmom}
    \sigma_{ij,i}=0\quad\text{where}\quad \sigma_{ij}=\sigma_{ji},
\end{equation}
with the \textsc{Cauchy} stress tensor $\sigma_{ij}$, subject to the boundary conditions
\begin{equation}
    \begin{aligned}
    \sigma_{ij} \, n_i&=\bar{t}_j \quad\text{on}\quad \partial\Omega_{t}\quad \text{and}\\
    u_i&=\bar{u}_i\quad\text{on}\quad \partial\Omega_{u}\coma
    \end{aligned}
\end{equation}
where the boundary $\partial\Omega=\partial\Omega_{t}\cup\partial\Omega_{u}$ has been decomposed into a part $\partial\Omega_{t}$ with natural boundary conditions and a part $\partial\Omega_{u}$ with essential boundary conditions, and $\emptyset=\partial\Omega_{t}\cap\partial\Omega_{u}$.
The symmetry of the stress tensor follows from the angular momentum balance. The stress is energetically conjugate to the strain rate $\dot{\varepsilon}_{ij}$, with the strain defined as
\begin{equation}
    \varepsilon_{ij}=\frac{1}{2}\left(u_{i,j}+u_{j,i}\right)\coma
\end{equation}
in a geometrically linear setting, and the displacement $u_i$. 

Following Borden~\cite[p. 63 ff.]{borden_isogeometric_2012} or Kuhn~\cite[p. 41 ff.]{kuhn_numerical_2013}, micro forces are introduced as energetically conjugate to the phase-field rate $\dot{c}$. The according conservation equation in the local form reads
\begin{equation}
\label{eq:micro forcebalnance}
    \xi_{i,i}+\pi+\kappa=0\coma
\end{equation}
where $\xi_i$ is the micro force traction and $\pi$ and $\kappa$ are internal and external volume forces, respectively. After some manipulations and consideration of the first and second laws of thermodynamics, the \textsc{Clausius-Duhem} inequality
\begin{equation}
    \label{eq:cdu}
    \left[\sigma_{ij}\left(\varepsilon_{kl},c\right)-\frac{\partial\psi}{\partial\varepsilon_{ij}}\right]\dot{\varepsilon}_{ij}+\left[\xi_i\left(c,c_{,j},\dot{c}\right)-\frac{\partial\psi}{\partial c_{,i}}\right]\dot{c}_{,i}-\left[\pi\left(c,c_{,j},\dot{c}\right)+\frac{\partial\psi}{\partial c}\right]\dot{c}\geq 0
\end{equation}
is derived. Here, the argument of Gurtin~\cite{gurtin_generalized_1996} has been employed, that the free energy density $\psi\left(\varepsilon_{ij},c,c_{,k}\right)$ must not be a function of $\dot{c}$. The dependencies of $\psi$ are dropped above and below for the sake of readability. Furthermore, $\dot{\varepsilon}_{ij}$ and $\dot{c}_{,i}$ appear linearly: If Equation~\eqref{eq:cdu} shall hold for any admissible $\dot{\varepsilon}_{ij}$ and $\dot{c}_{,i}$, the constitutive relations
\begin{align}
    \label{eq:stress}
    \sigma_{ij}(\varepsilon_{kl},c)&=\frac{\partial\psi}{\partial\varepsilon_{ij}}\quad\text{and}\\
    \label{eq:xi}
    \xi_i\left(c,c_{,j},\dot{c}\right)&=\frac{\partial\psi}{\partial c_{,i}}
\end{align}
can be deduced. 
Following the argument of Gurtin~\cite{gurtin_generalized_1996} or Kuhn~\cite[p. 41 ff.]{kuhn_numerical_2013}, the last term can be satisfied if
\begin{equation}
    \label{eq:pi}
    \pi\left(c,c_{,j},\dot{c}\right)=-\eta_\text{f} \, \dot{c}-\frac{\partial\psi}{\partial c}\coma
\end{equation}
where $\eta_\text{f}\geq 0$ serves as a kinetic fracture parameter or viscosity. Inserting Equations~\eqref{eq:xi} and~\eqref{eq:pi} into Equation~\eqref{eq:micro forcebalnance}, a \textsc{Ginzburg-Landau}-type equation
\begin{equation}
    \label{eq:pfevolution}
    \eta_\text{f} \, \dot{c}=\left(\frac{\partial\psi}{\partial c_{,i}}\right)_{,i}-\frac{\partial\psi}{\partial c}
\end{equation}
is obtained. As the phase-field $c$ is not influenced by any external quantity directly, a zero external micro volume force~$\kappa=0$ and the homogeneous boundary condition
\begin{equation}
    \label{eq:pfbc}
    \xi_i\left(c,c_{,j},\dot{c}\right)n_i=0 \quad\text{on}\quad\partial\Omega
\end{equation}
for Equation~\eqref{eq:pfevolution} are defined. The constitutive ansatz for $\psi$ is discussed in the next section.

\subsection{Constitutive modeling of material response}
\label{sec:constitMod}
The constitutive modeling approach follows an additive split of the total free \textsc{Helmholtz} energy
\begin{equation}
    \Psi=\int\limits_\Omega \psi\dv=\int\limits_\Omega \psi^\text{el}\dv + \int\limits_\Omega \psi^\ellc\dv
\end{equation}
into an elastic $\psi^\text{el}$ and a phase-field $\psi^\ellc$ contribution. For the elastic term, the widely used tensile split~\cite{miehe_thermodynamically_2010}
\begin{align}
    \psi^\text{el}=g(c) \, \psi^\text{el}_{0,+} + \psi^\text{el}_{0,-}\quad&\text{with}\quad \psi^\text{el}_0=\frac{E(x_l)\,\nu}{2(1-2\nu)(1+\nu)}\left(\varepsilon_{kk}\right)^2+ \frac{E(x_l)}{2(1+\nu)}\varepsilon_{ij} \varepsilon_{ij}
\label{eq:tenssplit}
\end{align}
has been adopted. Only the tensile part $\psi^\text{el}_{0,+}$ is degraded using the degradation function $g(c)=c^2+\eta$ to prevent crack forming under pressure. A small residual stiffness $\eta=10^{-6}$ is maintained for the fully degraded ($c=0$) state. The \textsc{Young}'s modulus $E(x_l)$ may exhibit a spatial dependence, cf. Section~\ref{sec:youngtanh}, while the \textsc{Poisson} ratio $\nu$ is assumed to be constant in the remainder of this paper. The tensile split does not fully degrade the material under shear~\cite{ambati_review_2015}. A remedy to this issue is the physically based split~\cite{steinke_phase-field_2018}. Possible impacts on the results are discussed below. 

The energy apparently stored within the phase-field contribution takes the form
\begin{equation}
    \psi^\ellc =  \frac{\mathcal{G}_\text{c}(x_l)}{4\ellc}\left[ \left(1-c\right)^2 +4\ellc^2 c_{,i}c_{,i}\right] = \mathcal{G}_\text{c}(x_l)\, \gamma^{\ellc}\coma
\end{equation}
which stems from the energetic criterion of Griffith~\cite{griffith_phenomena_1921}. It is noted, that the fracture toughness $\mathcal{G}_\text{c}(x_l)$ may exhibit a spatial dependence, see Section~\ref{sec:gcreduction}.

With the constitutive model at hand, it is possible to deduce more specific expressions for the stress
\begin{equation}
    \sigma_{ij}=g(c)\frac{\partial\psi^\text{el}_{0,+}}{\partial\varepsilon_{ij}}+\frac{\partial\psi^\text{el}_{0,-}}{\partial\varepsilon_{ij}}\coma
\end{equation}
the evolution equation for the phase-field
\begin{equation}
\label{eq:pfevolution2}
    \eta_\text{f} \, \dot{c}=\frac{\mathcal{G}_\text{c}(x_l)}{2\ellc}+2\ellc\left(\mathcal{G}_\text{c}(x_l)\, c_{,i}\right)_{,i}-c\left(\frac{\mathcal{G}_\text{c}(x_l)}{2\ellc}+2\psi^\text{el}_{0,+}\right)    
\end{equation}
and the corresponding boundary condition
\begin{equation}
\label{eq:pfbc2}
    c_{,i} \, n_i=0 \quad\text{on}\quad\partial\Omega
\end{equation}
from Equations~\eqref{eq:stress}, \eqref{eq:pfevolution} and~\eqref{eq:pfbc}.

In order to prevent existing cracks from healing, an irreversibility constraint has to be imposed. There are two widespread approaches, the \textit{damage-like} and \textit{fracture-like} irreversibility condition. The former one interprets the phase-field as damage variable and requests $\dot{c}\leq0$ in every material point. This can be achieved by introducing a history variable~\cite{miehe_phase_2010}. The latter approach allows for local reversibility and does not constraint the phase-field before it reaches a critical threshold close to zero~\cite{kuhn_continuum_2010}. Then, a \textsc{Dirichlet} boundary condition $c=0$ is set at the corresponding location. The advantage of the \textit{fracture-like} constraint is, that the dissipated energy associated with the crack surface is not overestimated~\cite{linse_convergence_2017}. In this contribution, the latter approach is chosen with a threshold of $c_\text{th}=0.03$. A study for different values of $c_\text{th}$ did not reveal any significant differences between the simulation results obtained for this value and those for higher $c_\text{th}$.
Additionally, no further influence of a varying characteristic element size from five to ten times smaller than $\ellc$ was observed for a fixed value of $c_\text{th}=0.03$.

\subsection{Weak form and finite element implementation}
\label{sec:modelderiv}
The weak forms of the partial differential equations~\eqref{eq:linmom} and~\eqref{eq:pfevolution2} are obtained as follows. Both equations are multiplied with test functions $\delta u_j$ and $\delta c$, and integrated over the whole domain. Integration by parts and making use of the divergence theorem yields
\begin{align}
    0&=\int\limits_\Omega \sigma_{ij}\,\delta u_{j,i}\dv-\int\limits_{\partial\Omega_t}\bar{t}_j \delta u_j\da \quad\text{and}\\
    \label{eq:weakpf}
    0&=\int\limits_\Omega \left[ \frac{\mathcal{G}_\text{c}(x_l)}{2\ellc}-c\left(\frac{\mathcal{G}_\text{c}(x_l)}{2\ellc}+2\psi^\text{el}_{0,+}\right)-  \eta_\text{f} \, \dot{c}\right]\delta c - 2\ellc\mathcal{G}_\text{c}(x_l)\, c_{,i} \delta c_{,i}  \dv+\int\limits_{\partial\Omega}\underbrace{ 2\ellc\mathcal{G}_\text{c}(x_l)\, c_{,i} n_i\, \delta c}_{=0,\text{ cf. Eq.~\eqref{eq:pfbc2}}}\da\point
\end{align}
A time discrete form is obtained, by replacing the phase-field rate in Equation~\eqref{eq:weakpf} using an \textsc{Euler} backward scheme
\begin{equation}
    \dot{c}\approx \frac{c- {^nc}}{\Delta t}\coma
\end{equation}
where $^nc$ is the converged phase-field value of the previous increment and $\Delta t$ is the time step.

The open-source finite element package FEniCS and the numerics library PETSc~\cite{arge_efficient_1997} allow for an efficient parallelized solution of differential equations. An important ingredient of the framework is the so-called \textit{Unified Form Language} (UFL)~\cite{alnaes_unified_2014}, a python-based language for mathematical expressions. The implementation is carried out using the Python interface of FEniCS by stating the weak form and all necessary constitutive relations using UFL.
From that, a parallelized code for the solution of the finite element system is automatically generated~\cite{kirby_compiler_2006,logg_dolfin:_2010}.
Formulating the model equations in a UFL-conforming manner, special attention has to be paid to the necessary spectral decomposition of the strain tensor due to the tensile split.
The key aspects of the implementation are explained in \ref{sec:impl}. Furthermore, the complete formulation of the spectral decomposition as well as the implementation of the constitutive relations and the weak forms of the governing equations are provided as supplementary material.\footnote{Please cite this article if reused in any form.}

The resulting non-linear, time-discrete equations are solved using a fully coupled, monolithic approach.
In order to avoid convergence problems, the \textit{backtracking} line search algorithm provided by PETSc is enabled.
Furthermore, a heuristic adaptive time-stepping scheme is employed, which reduces the time step, if the \textsc{Newton-Raphson} scheme reaches no convergence within 70 iterations and increases the time step if convergence is reached within four iterations.

Along the interface and the crack path, the mesh is refined such that the characteristic element length is five to eight times smaller than $\ellc$. 
Spatial convergence has been verified.

\subsection{Interface modeling in the context of regularized heterogeneities}
\begin{figure}[t]
    \centering
    \subfloat[Intact and broken interface according to LEFM~\cite{he_crack_1989}]{
        \label{fig:discif}
        \includegraphics[trim={0cm 0cm 8cm 0cm},clip]{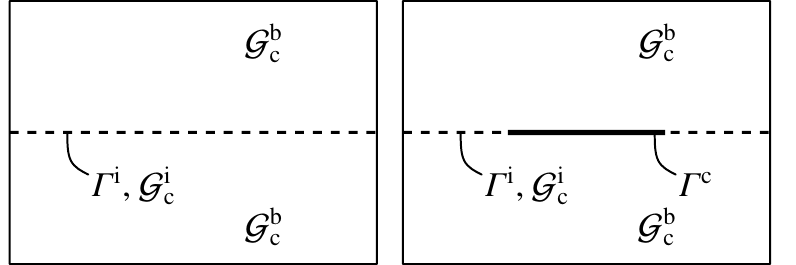}
    }
    \subfloat[Intact and broken interface in a regularized setting]{
        \label{fig:diffif}
        \includegraphics[trim={8.2cm 0cm 0cm 0cm},clip]{fig02ab.pdf}
    }
    \caption{Intact and partially cracked interface for a two-dimensional specimen: The interface $\gammai$ in LEFM is depicted by a line with a fracture toughness $\gci$ different from the bulk material fracture toughness $\gcb$, \textbf{(a)} left picture. Failure leads to a sharp crack $\gammac$, \textbf{(a)} right picture. For a regularized interface $\gammali$ according to~\cite{hansen-dorr_phase-field_2019}, the zone, where the fracture toughness deviates from $\gcb$, has a finite width. The regularization is schematically depicted by the grey hatched area, \textbf{(b)} left picture. The parameter $\elli$ measures the width along the direction of the signed distance $d$. Failure leads to a regularized, phase-field crack $\gammalc$ with characteristic length $\ellc$, \textbf{(b)} right picture.}
    \label{fig:discvsdiffif}
\end{figure}

In the context of LEFM, interfaces are mostly introduced as infinitesimal layers of $D-1$ physical dimensions, where $D$ is the dimension of the considered domain separated by the interface. Such a description is also chosen in the work of He and Hutchinson~\cite{he_crack_1989,he_kinking_1989}, who investigated crack deflection and branching at interfaces. Surrounded by two, possibly dissimilar, bulk materials $i=1,2$ with elastic $E_i,\nu_i$ and fracture $\gcb$ material parameters, the interface $\gammai$ is only assigned an interface fracture toughness $\gci$, cf. Figure~\ref{fig:discif}. A crack $\gammac$ emerging along the interface has $D-1$ physical dimensions, too.

Hansen-Dörr et al.~\cite{hansen-dorr_numerical_2017,hansen-dorr_phase-field_2018,hansen-dorr_phase-field_2019} have introduced a regularized interface model which allows for non-conforming interfaces within a regular mesh. In analogy to classical phase-field models, the interface is regularized and defined as a subdomain $\gammali$ of $D$ physical dimensions, which separates at least two other subdomains of materials with possibly dissimilar elastic properties. The interface mid-surface is identical to the discrete interface $\gammai$. The difference of the interface $\gammali$ to other sub-structures is, that one physical dimension is considerably smaller than the smallest characteristic lengths of every other subdomain (except from other interfaces). This property is called \textit{narrow} and is quantified by introducing the length scale $\elli$, which measures the width in the direction of the signed distance $d$, cf. Figure~\ref{fig:diffif}. It is further assumed, that the elastic energy stored within such a regularized interface is negligibly small compared to non-interfacial subdomains. Thus, in analogy to the LEFM description, the interface is not explicitly assigned exclusive elastic parameters but also values depending on the surrounding bulk materials. Despite this simplification the fracture toughness $\gci$ is still relevant and can significantly influence the macroscopic cracking behavior of a structure, even if the interface width is macroscopically not recognizable. A crack along the interface is regularized, too, and becomes a phase-field crack $\gammalc$ with the characteristic length $\ellc$.

\subsubsection{Incorporation of the interface by means of a fracture toughness reduction}
\label{sec:gcreduction}

\begin{figure}[t]
    \centering
    \subfloat[Regularization of the fracture toughness]{
        \label{fig:gcregul}
        \begin{minipage}[t]{0.4\textwidth}
        \centering
        \footnotesize
        \includegraphics{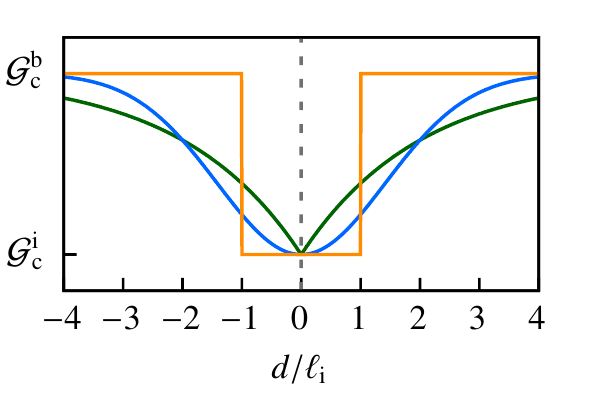}
	    \begin{tabular}{l|l|l}
		    \legline{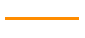} $\rheavi$ & \legline{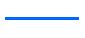} $\rgauss$ & \legline{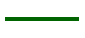} $\rexpo$ \\   
	    \end{tabular}
        \end{minipage}
    }
    \subfloat[Regularization of the \textsc{Young}'s modulus]{
        \label{fig:youngsregul}
        \begin{minipage}[t]{0.4\textwidth}
        \centering
        \footnotesize
        \includegraphics{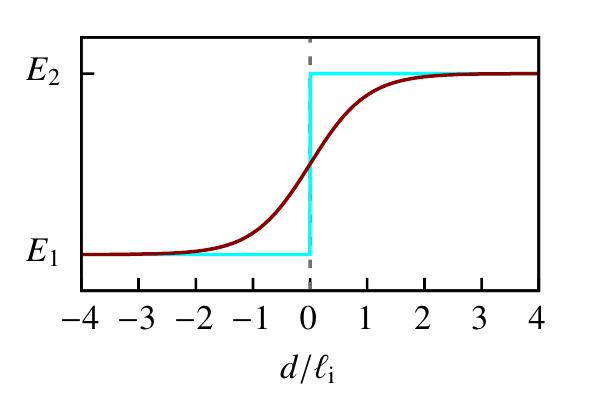}
	    \begin{tabular}{l|l}
		    \legline{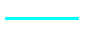} $\reheavi$ & \legline{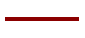} $\rtan$ \\   
	    \end{tabular}
        \end{minipage}
    }
    \caption{Regularization curves for the material parameters: \textbf{(a)} depicts the fracture toughness distribution for the different functions $\rheavi$, $\rgauss$ and $\rexpo$. The signed distance $d(x,y)$ measures the shortest distance from any point within the domain to the interface midline. \textbf{(b)} depicts the spatial distribution of the \textsc{Young}'s modulus $\reheavi$ and $\rtan$. In both plots, the location of the interface midline $d/\elli=0$ is highlighted by a dashed~\legline{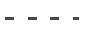} line.} 
    \label{fig:paramregul}
\end{figure}

An interface, which is schematically depicted in Figure~\ref{fig:diffif}, can be described using a \textsc{Heaviside}-like function for the fracture toughness
\begin{equation}
\label{eq:rheavi}
\rheavi(d,\gcb,\gci)=\begin{cases}
\gcb& \text{for } \vert d \vert > \elli\\
\gci& \text{for } \vert d \vert \leq \elli 
\end{cases}    
\end{equation} 
within the whole domain. The spatial dependence of $\mathcal{G}_\text{c}^\text{H}$ is implicitly incorporated by the signed distance $d$, which measures the shortest distance from every point to the interface midline. This description was used by Hansen-Dörr et al.~\cite{hansen-dorr_phase-field_2019}, where it was shown that non-conforming interfaces can be described, despite the jump of the fracture toughness, which may occur within an element. However, the sharp switch between two fracture toughness values negatively influenced the convergence of the numerical solver. Besides the \textsc{Heaviside}-like description, an exponentially-shaped
\begin{equation}
    \rexpo(d,\gcb,\gci)=\gcb-\left( \gcb-\gci \right) \,\exp\left[-\frac{\left| \, d \, \right|}{2\elli}\right]
\end{equation}
and a \textsc{Gaussian}-like
\begin{equation}
    \rgauss(d,\gcb,\gci)=\gcb-\left( \gcb-\gci \right) \,\exp\left[-\left(\frac{d}{2\elli}\right)^2\right]
\end{equation}
function are investigated. All three regularization functions\footnote{In the context of the fracture phase-field, the authors understand the regularization of the interface as increase of its dimension, i.e. from $D-1$ in LEFM to $D$ as in the present model within a $D$-dimensional domain.} are depicted in Figure~\ref{fig:gcregul}. The comparison clearly reveals, that the regularizations $\rgauss$ and $\rexpo$ introduce a transition zone, which is larger than $\elli$. However, in the context of the regularized phase-field model, the length $\elli$ can still be identified as characteristic interface width in analogy to the characteristic crack length $\ellc$. Additionally, the differentiation of the bulk material and the interface is softened by introducing a continuous regularization: The interface is no longer an additional material stripe, which can clearly be identified, but a diffuse region.\footnote{Referring to Figure~\ref{fig:diffif} this means, that the grey stripe has to be understood as a symbol for the interface regularization and not a sharp differentiation from the surrounding bulk material in the context of $\rgauss$ and $\rexpo$.} 

\subsubsection{Incorporation of elastic heterogeneities near the interface}
\label{sec:youngtanh}
In the proposed model, the interface formally has the same number of material parameters as the surrounding bulk material. In earlier investigations with the model~\cite{hansen-dorr_phase-field_2018,hansen-dorr_combined_2019}, only elastically homogeneous cases were investigated. The elastic constants of the two bulk materials were also applied within the interface even if it was in principle possible to consider completely different elastic constants, in a similar fashion as in Equation~\eqref{eq:rheavi}. However, as outlined above, the deformation energy of the interface is assumed to be negligibly small compared to the bulk materials' deformation energy. This description is consistent with the assumption of analytic LEFM calculations~\cite{he_crack_1989}, where the $D-1$-dimensional interface neither has elastic properties.

In this work, the interface is interpreted as a transition zone with respect to the elastic material parameters. In principle, the transition could be modeled using a whole variety of different functions. In this work, only the \textsc{Young}'s modulus is allowed to vary, while the \textsc{Poisson} ratio $\nu$ is assumed constant for the present investigations. The modulus follows a hyperbolic tangent-like shape
\begin{equation}
\label{eq:rtan}
\rtan(d)=\frac{E_2-E_1}{2}\left(\tanh\left[\frac{d}{\elli} \right]+1\right)+E_1\point
\end{equation}
For some investigations, the smooth \textsc{Young}'s modulus transition is compared to the sharp limit, which reads
\begin{equation}
	\reheavi(d)=\begin{cases}
		E_2& \text{for } d > 0\\
		E_1& \text{for } d \leq 0 \point
	\end{cases}   
\end{equation}
It is further noted, that the length scales for the fracture toughness and \textsc{Young}'s modulus regularizations are both chosen in dependence of $\elli$. Generally, these values could be independent from each other. However, it makes sense to choose them in the same order of magnitude because they govern the spatial discretization, too. The functions $\reheavi$ and $\rtan$ are depicted in Figure~\ref{fig:youngsregul} for $E_2>E_1$, but not restricted to this condition.

The concept of regularizing jumps in the elastic constants according to Equation~\eqref{eq:rtan} is not new and has widely been used in literature, for example Schneider et al.~\cite{schneider_phase-field_2015}, Mosler et al.~\cite{mosler_novel_2014} and Kiefer et al.~\cite{kiefer_numerical_2017}, where a phase-field model is used to describe phase transitions. Equation~\eqref{eq:rtan} can thus be understood as a static phase-field. The assumption of such a transition might however lead to unwanted behavior in the vicinity of the interface as just stated. Physically not reasonable effects like an exaggerated, interfacial energy~\cite{schneider_phase-field_2015} or a violation of the mechanical jump conditions~\cite{schneider_phase-field_2015,mosler_novel_2014} may result. A possible solution is the so-called \textit{partial rank-I relaxation}, which accounts for the mechanical equilibrium in every material point. 

In this contribution, no such approach is implemented at the cost of possible inaccuracies near the interface. The reason is that the combination of the tensile split introduced in Section~\ref{sec:modelderiv} and a \textit{partial rank-I relaxation} is non-trivial. The error which is made, is quantified below by a comparison to results obtained with a sharp, mesh-conforming elastic jump $\reheavi$.

\subsection{Configurational forces and link to energy release rate}
\label{sec:confforces}
The scope of this work is not only the qualitative analysis of various crack patterns in heterogeneous materials but also the quantification of the so-called crack driving forces, which lead to the aforementioned and yield a deeper understanding of why and when branching and deflection occur. Rice~\cite{rice_path_1968} and Cherepanov~\cite{cherepanov_crack_1967} developed the concept of a path independent integral, the $\mathcal{J}$-integral. The evaluation serves as an alternative way to calculate the energy release rate $\mathcal{G}$ in LEFM. Later, the $\mathcal{J}$-integral was generalized for multidimensional analyses, cf.~\cite{delorenzi_energy_1982,ohtsuka_generalized_1985,kuna_numerische_2008}. For the specific application within a coupled mechanical crack phase-field framework, Kuhn and Müller~\cite{kuhn_continuum_2010,kuhn_numerical_2013,kuhn_discussion_2016} introduced a generalized configurational force balance, which is closely related to the generalized $\mathcal{J}$-integral, to account for heterogeneities within the material in the determination of the energy release rate. The configurational force balance of the deformation energy
\begin{equation}
\label{eq:confforcebal}
\Sigma_{ij,i}^\text{def}+\tilde{g}_j^\text{modE}+\tilde{g}_j^\text{dis}+\tilde{g}_j^\text{tip}=0
\end{equation}
enables the computation of the crack driving forces. The individual contributions break down as follows. The deformation energy contribution
\begin{equation}
\Sigma_{ij}^\text{def}=\psi^\text{el}\delta_{ij}-u_{k,i}\sigma_{kj}
\end{equation}
is identical to the integrand of the generalized $\mathcal{J}$-integral, cf.~\cite{kuna_numerische_2008}. The contribution of varying elastic material parameters manifests itself in
\begin{equation}
\tilde{g}_i^\text{modE}=\left.-\frac{\partial \psi^\text{el}}{\partial x_i}\right|_\text{explicit}\coma
\end{equation}
accounting for the explicit spatial dependence -- in this work due to a varying \textsc{Young}'s modulus. The influence of the dissipative, viscous term is incorporated in
\begin{equation}
\tilde{g}_i^\text{dis}=\eta_\text{f}\, \dot{c} \,c_{,i} \point
\end{equation}
Following Kuhn~\cite{kuhn_discussion_2016}, the crack driving force 
\begin{equation}
\label{eq:jint}
\jt_i=-\int\limits_\mathcal{C}\tilde{g}_i^\text{tip}\dv
\end{equation}
can be calculated by integrating over a circular volume of unit thickness
\begin{equation}
    \mathcal{C}=\left\lbrace x,y \,\middle\vert\, (x-x_\text{tip})^2+(y-y_\text{tip})^2\leq r^2 \right\rbrace
\end{equation}
with radius $r$ centered at the crack tip $[x_\text{tip}\,\, y_\text{tip}]\T$. The numerical evaluation is based on the weak form of Equation~\eqref{eq:confforcebal} to avoid the calculation of the divergence, cf.~\cite{kuhn_discussion_2016}. In analogy to LEFM, $\jt_i$ can be named generalized $\mathcal{J}$-integral. In contrast to the classical $\mathcal{J}$-integral it can be applied to locally heterogeneous structures. For the sake of simplicity, only the most important implications and relations have been mentioned here. 

In general, the choice of $\mathcal{C}$ may influence the results of Equation~\eqref{eq:jint}, especially, when the radius is chosen too small or larger than the simulation domain. In order to arrive at an appropriate decision, the integral is evaluated for many different radii and the results are compared concerning converged integral component values. These components $\jt_i$, $i=x,y$, now reflect the energy release rates with respect to the chosen coordinate frame, which is why the term \textit{energy release rate} is used in the remainder of this paper for the discussion of individual components of $\jt_i$. In this work, a comparison of different radii revealed $r=\SI{0.35}{\milli\meter}$ to be a good choice. 

\subsection{One-dimensional phase-field profiles at interfaces}
\label{sec:optprof}
Hansen-Dörr et al.~\cite{hansen-dorr_numerical_2017,hansen-dorr_phase-field_2018,hansen-dorr_phase-field_2019} have observed that for homogeneous elastic properties and a fracture toughness variation according to $\rheavi$, a straight mode-I crack does not propagate along the interface for a critical energy release rate
equal to the fracture toughness of the interface $\gci$, but a higher value $\gcia$ between $\gci$ and $\gcb$. The exact value of the actual fracture toughness of the interface $\gcia$ depends on the ratios $\gcb/\gci$, $\elli/\ellc$ and the exact function $\rheavi$, $\rgauss$ or $\rexpo$ which is used. In the following, one-dimensional considerations are presented which illustrate this phenomenon.

\begin{figure}[t]
	\centering
	\subfloat[Analytical phase-field profile for heterogeneous $\mathcal{G}_\text{c}$]{
		\label{fig:anaprof_1D}
		\begin{minipage}[t]{0.4\textwidth}
			\centering
			\footnotesize
			\includegraphics{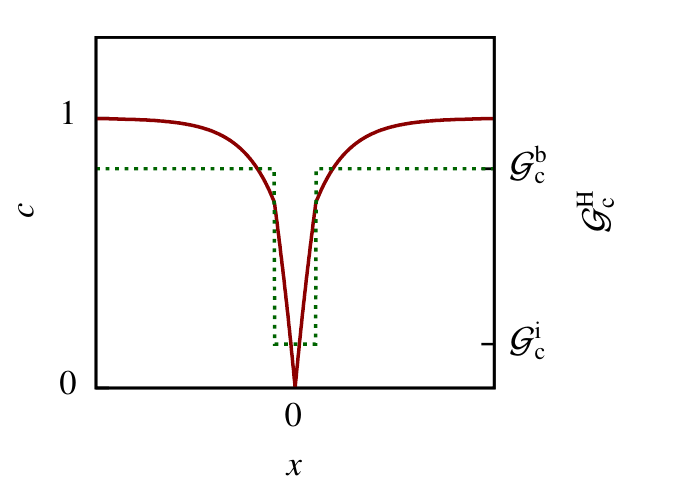}
			\begin{tabular}{l|l}
				\legline{3} $c$ & \legline{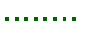} $\rheavi$ \\   
			\end{tabular}
		\end{minipage}
	}
	\subfloat[Analytical compensation curves for three ratios $\elli/\ellc$]{
		\label{fig:anacomp_1D}
		\begin{minipage}[t]{0.6\textwidth}
			\centering
			\footnotesize
			\includegraphics{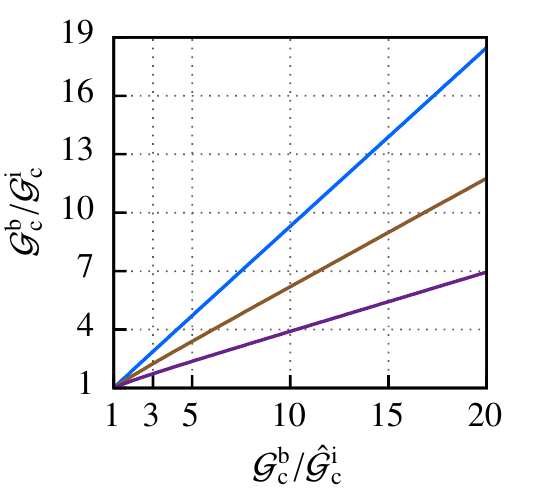}
			\begin{tabular}{l|l|l}
				\legline{5} $\elli/\ellc=3.125$ & \legline{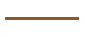} $\elli/\ellc=1.25$ & \legline{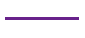} $\elli/\ellc=0.625$\\  
			\end{tabular}
		\end{minipage}
	}
	\caption{In \textbf{(a)}, the analytical one-dimensional phase-field
		profile for a heterogeneous fracture toughness is depicted. It exhibits a kink, where the fracture toughness is discontinuous. This yields an overestimation of the interface fracture toughness, which is accounted for by the compensated interface fracture toughness $\gcih$ in \textbf{(b)}.}
	\label{fig:anacomp}
\end{figure}
The analytical solution for the phase-field profile in the one-dimensional case~\eqref{eq:1dregcrack} is recovered for a homogeneous fracture toughness. Suppose a one-dimensional rod with $x\in (-\infty,\infty)$ of cross-sectional area $A$ is cracked at $x=\SI{0}{\milli\meter}$ where an interface is located, which is depicted by $\rheavi(x-\elli,\gcb,\gci)$, cf. Equation~\eqref{eq:rheavi}. The phase-field profile which forms under these circumstances can be calculated by considering the spatially dependent fracture toughness in the minimization of the functional
\begin{equation}
    \label{eq:defCSD1d_heavi}
    I\left[c,c^\prime\right]=\int\limits_{\Omega} \frac{\rheavi}{4\ellc}\left[ \left(1-c\right)^2 +4\ellc^2\left( c^\prime\right)^2\right]\dv\rightarrow \text{min}\point
\end{equation}
The solution for $x\in [0,\infty)$ reads
\begin{equation}
\label{eq:solHet}
    c(x)=\begin{cases}
		1+O\,\exp[-x/(2\ellc)]& \text{for } x > \elli\\
		1+P\,\exp[x/(2\ellc)]+Q\,\exp[-x/(2\ellc)] & \text{for } x \leq \elli
	\end{cases}   
\end{equation}
with the integration constants
\begin{equation}
\label{eq:1Dconst}
    O=  \frac{-2}{\left(1-\gcb/\gci\right)\exp[-\elli/\ellc]+\gcb/\gci+1}\,,\quad   P=  O\,\frac{1-\gcb/\gci}{2}\exp[-\elli/\ellc]\quad\text{and}\quad
    Q=  O\,\frac{1+\gcb/\gci}{2}\point
\end{equation}
Due to symmetry, the solution for $x\in (-\infty,0)$ can be obtained by mirroring at $x=0$. The phase-field profile for a cracked interface with $\gcb/\gci = 3$ and $\elli/\ellc = 1.25$ is depicted in Figure~\ref{fig:anaprof_1D} and clearly exhibits a kink at $\vert x \vert = \elli$ where $\gci$ switches to $\gcb$. We refer to the phenomenon of an altered phase-field profile as length scale interaction of the crack and the interface lengths. Inserting the solution~\eqref{eq:solHet} into the crack surface density and integrating over the domain gives a different value of the total crack surface with respect to the case of a homogeneous fracture toughness. The same applies to the surface energy of the crack, which is overestimated depending on the ratios of the length scales and fracture toughnesses. This is inconvenient because the crack energy is not independent from internal length scales. In other words, assuming an interface crack, the actual interface fracture toughness $\gcia$ for the one-dimensional example is not equal to the specified interface fracture toughness $\gci$  in Figure~\ref{fig:anaprof_1D}. After some straightforward manipulation of Equation~\eqref{eq:defCSD1d_heavi}, which describes the total crack energy, and making use the analytical solution~\eqref{eq:solHet}, one obtains
\begin{equation}
\label{eq:comp1D}
\frac{\gcb}{\gcia} = \frac{\gcb/\gci}{P^2 [\exp(\elli/\ellc)-1]  -Q^2 [\exp(-\elli/\ellc)-1]+O^2[ \gcb/\gci\exp(-\elli/\ellc)] }\point
\end{equation}
The compensation of the length scale interaction to achieve length scale independent results can be motivated with Equation~\eqref{eq:comp1D} by requiring $\gcia=\gci$. Consequently, a compensated interface fracture toughness $\gcih$ has to be used for the interface description $\rheavi(x-\elli,\gcb,\gcih)$, which results in exchanging $\gci$ with $\gcih$ on the right hand-side of Equation~\eqref{eq:comp1D}. A graphical representation for different length scale ratios, where $\gcia=\gci$ is required, is given in Figure~\ref{fig:anacomp_1D}. It becomes apparent from Equations~\eqref{eq:1Dconst} and~\eqref{eq:comp1D}, that the compensation is only depending on ratios $\gcb/\gci$ and $\elli/\ellc$, and not on absolute values.

\subsection{Numerical two-dimensional compensation approach for length scale interaction}
\label{sec:compappr}
\begin{figure}[t]
	\centering
	\subfloat[\textsc{Heaviside}-like description]{
		\includegraphics{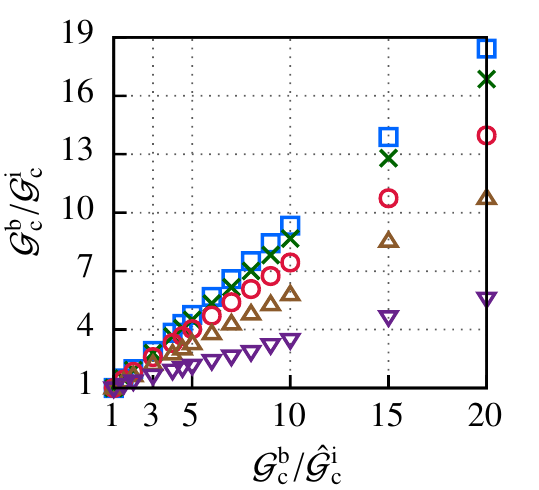}
		\label{fig:compensation_h}
	}
	\subfloat[\textsc{Gaussian}-like regularization]{
		\includegraphics{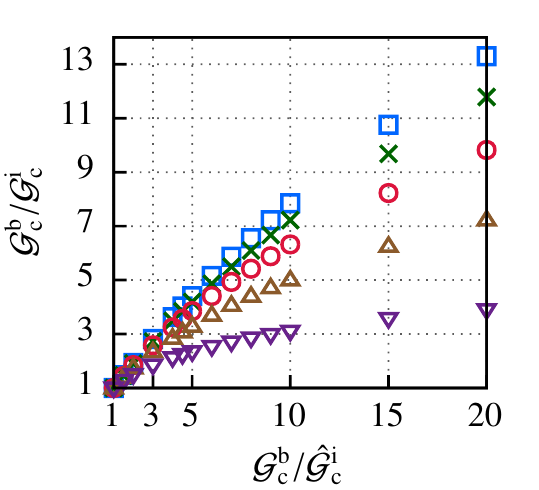}
		\label{fig:compensation_g}
	}
	\subfloat[Exponential regularization]{
		\includegraphics{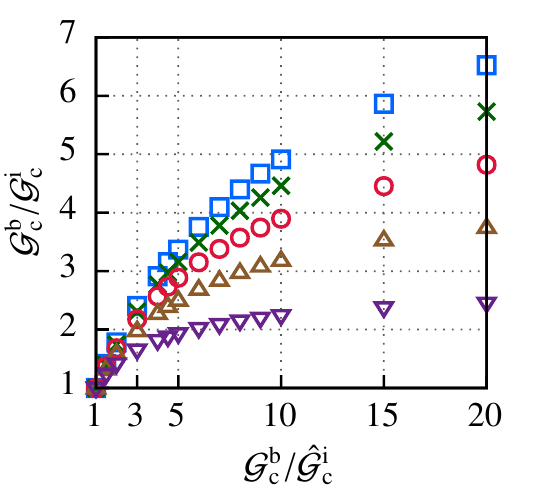}
		\label{fig:compensation_e}
	}
	\\
	\vspace{0.3cm}
	\footnotesize 
	\begin{tabular}{l|l|l|l|l}
		\legpoint{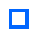} $\elli/\ellc=3.125$ & \legpoint{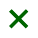} $\elli/\ellc=2.5$ & \legpoint{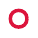} $\elli/\ellc=1.875$ & \legpoint{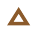} $\elli/\ellc=1.25$ & \legpoint{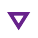} $\elli/\ellc=0.625$\\   
	\end{tabular}
	\caption{Compensation of the length scale interaction of $\elli$ and $\ellc$: Depending on the regularization \textbf{(a)}~--~\textbf{(c)} and the ratios $\gcb/\gci$ and $\elli/\ellc$, a compensated fracture toughness $\hat{\mathcal{G}}_\text{c}^\text{i}$ can be determined, which accounts for the bulk material influence and leads to an interfacial crack resistance of $\gci$. 
	}
	\label{fig:compensation}
\end{figure}

For a more complex fracture toughness heterogeneity $\rexpo$ or $\rgauss$, an analytical solution does not seem feasible. Furthermore, for actual crack propagation simulations, the coupled field problem, cf. Section~\ref{sec:modelderiv}, is solved, which adds a source term to the phase-field equation. Applying the analytical compensation introduces inaccuracies. Therefore, the corresponding compensation plots are obtained numerically by a mode-I crack simulation along an interface. A parameter study for various $\gcb/\gcih$ and $\elli/\ellc$ similar to~\cite{hansen-dorr_phase-field_2019} has been carried out and the resulting ratios $\gcb/\gci$ are recorded making use of the configurational forces, cf. Section~\ref{sec:confforces}. The resulting compensation plots are depicted in Figure~\ref{fig:compensation}, where each symbol resembles one simulation. Comparing Figure~\ref{fig:compensation_h} to Figure~\ref{fig:anacomp_1D}, the inaccuracy of the one-dimensional approach for two-dimensional crack propagation becomes apparent. For small ratios $\elli/\ellc$ and rising ratios $\gcb/\gcih$, the discrepancy increases. Comparing Figures~\ref{fig:compensation}a--c, a crucial difference becomes clear. For the same length scale ratios, different values for the compensated interface fracture toughness have to be applied. In other words, if the interface regularization does not implement low fracture toughness values over a wide range across the interface, a comparably lower compensated interface fracture toughness has to balance the bulk fracture toughness influence. This effect increases from the \textsc{Heaviside}-like to the \textsc{Gaussian}-like to the exponential description and for the latter one, the saturation effect, which can be observed for every regularization, becomes the strongest, which is clearly a limitation. Its implications are discussed in Section~\ref{sec:numericresults}.

A further limitation of the presented compensation approach, which is applied in all ensuing simulations on crack branching and deflection, becomes apparent when looking at the compensation. It is based on a straight mode-I crack along the interface, which is not always an appropriate assumption. As soon as the crack is not aligned symmetrically, the compensation is distorted, which will become apparent later. The misalignment is also influenced by the crack tip tracking method, which is mentioned below. However, due to its simplicity and the diffuse character of both, the crack and the interface, this inaccuracy is accepted and quantified in the remainder. An extension to different bulk material fracture toughnesses left and right of the interface is possible but not subject of this work, because the reference results assume only one bulk material fracture toughness.

\section{Crack branching and deflection at interfaces}
\label{sec:numericresults}
The model presented above has been applied successfully to crack propagation along interfaces and it has been shown that the compensation is necessary for a quantitative comparison of crack driving quantities~\cite{hansen-dorr_phase-field_2019}. In this contribution, the setup is extended to a crack approaching an interface under a certain angle $\varphi_\text{i}$. Depending on the bulk material and interface properties, the crack branches, deflects or experiences no interfacial influence on its path. 

\begin{figure}[t]
    \centering
    \raisebox{-0.99cm}{\includegraphics{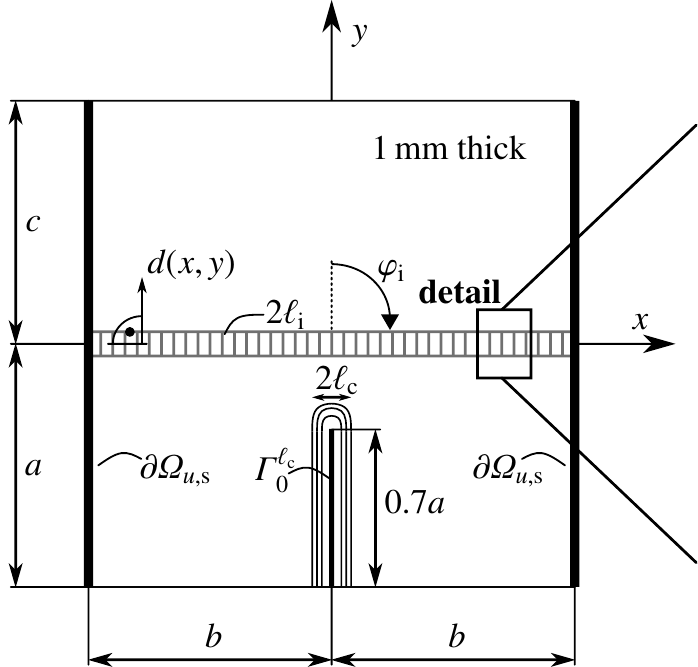}}\hspace{-0.4cm}
    \includegraphics[angle=90,origin=c,trim={0cm 0cm 0cm 0cm},clip]{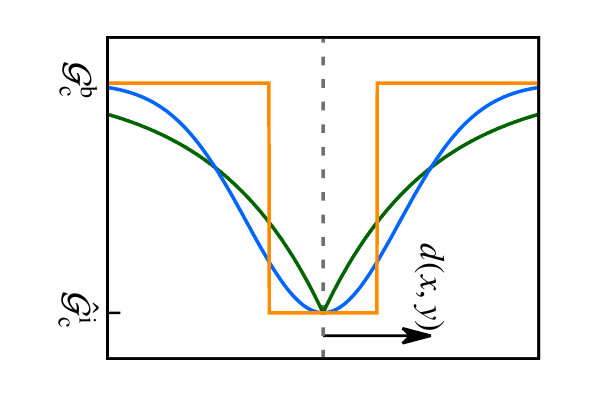}\hspace{-0.4cm}
    \includegraphics[angle=90,origin=c,trim={0cm 0cm 0cm 0cm},clip]{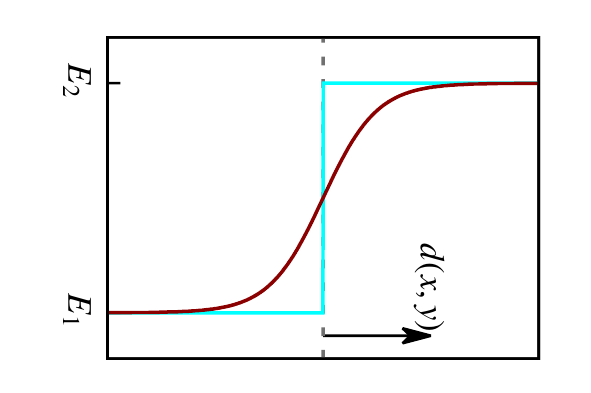}
    \caption{The general setup for the four different studies is sketched on the left. The regularized interface is schematically depicted by the grey hatched region, which may be inclined by a certain angle $\phii$. The material parameters are assigned according to the regularization functions depicted on the right, using values of $E_1$, $E_2$, $\hat{\mathcal{G}}_\text{c}^\text{i}$ and $\gcb$. The initial crack is depicted by $\gammalc_0$. \textsc{Dirichlet} boundary conditions are applied for the displacement along the bold marked side edges $\partial\Omega_{u,\text{s}}$ whereas homogeneous \textsc{Neumann} boundary conditions are applied on $\partial\varOmega\setminus \partial\varOmega_{u,\text{s}}$. 
    } 

    \label{fig:simsetup}
\end{figure}

Analytical considerations from He and Hutchinson~\cite{he_crack_1989} serve as a comparison. They investigated several crack-interface-configurations and have made predictions regarding the crack direction. Figure~\ref{fig:simsetup} captures all simulation setups which are dealt with below: Depending on the choice of the \textsc{Young}'s moduli $E_1$ and $E_2$, the bulk material and interface fracture toughnesses, $\gcb$ and $\gci$ respectively, and the interface inclination angle $\varphi_\text{i}$, four setups -- perpendicular or inclined interface, and homogeneous or heterogeneous elasticity -- serve as benchmarks. The \textsc{Poisson} ratio is not varied within the domain and the simulations have been conducted in a plane strain setting. For the investigations with a perpendicular interface, the domain measures are $a=b=c=\SI{1}{\milli\meter}$. For the inclined interface, different domain measures $a=b/2=c/3=\SI{1}{\milli\meter}$ are chosen to avoid that the interface passes through the corners of the specimen for the angles $\varphi_{\nv{i}}$ under consideration.

All interface regularization functions $\rheavi$, $\rgauss$ and $\rexpo$ are compared for the first study with a homogeneous \textsc{Young}'s modulus.
Any influence arising from the regularization for the elastic heterogeneity $\rtan$ is avoided in this way.

For the loading, the concept of the so-called \textit{surfing boundary condition} \cite{hossain_effective_2014} is exploited.
The key idea of this approach is to introduce a virtual crack tip with the time dependent position~$\bar{\ve{x}}$.
Here, a virtual tip moving along the $y$-axis,
\begin{equation}
    \bar{\ve{x}}
    = \begin{bmatrix}  \bar{x} \\ \bar{y}  \end{bmatrix}=
    \begin{bmatrix}  0 \\ v \cdot t + \bar{y}_0  \end{bmatrix}
   \coma
\end{equation}
is considered.
With respect to the virtual tip position, a displacement of hyperbolic tangent-like shape is applied on the side edges $\partial \varOmega_{u, \text{s}}$,
\begin{equation}
    \ve{u}^\nv{BC} =
  \frac{u_\nv{ref}^\nv{BC}}{2} \, \left( 1- \tanh \left[ \frac{y-\bar{y}}{d}\right] \right) \, \sign \left( x\right)
  \begin{bmatrix}  1 \\ 0  \end{bmatrix}
  \coma
 \label{eq:sbc-tanh}
\end{equation}
assuming $u_\text{ref}^\nv{BC} = \SI{7.5}{\micro\meter}$, $d=\SI{0.5}{\milli\meter}$, $v=\SI{0.3}{\milli m/s}$ and $\bar{y}_0=\SI{-2}{\milli m}$. 
In previous studies, this set of parameters proved to be suitable. Homogeneous \textsc{Neumann} boundary conditions are considered on the remaining part of the boundary $\partial\varOmega\setminus \partial\varOmega_{u,\text{s}}$.

All simulations presented within this paper are conducted assuming $\ellc = \SI{15}{\micro\meter}$.
The fracture toughness of the bulk material is set to the constant value $\gcb=\SI{2.7}{\newton \per\milli\meter}$ while $\gci$ is adapted according to the fracture toughness ratio $\gcb/\gci$ which is varied in order to study different fracture phenomena.
Similarly, $E_1=\SI{210}{\kilo\newton\per\milli\meter\squared}$ is considered and for the investigation of elastic heterogeneity, different values of $E_2$ are defined. A constant \textsc{Poisson} ratio $\nu=0.3$ is assumed.
The maximum and minimum time steps are $\Delta t_{\max} = \SI{8E-2}{\second}$ and $\Delta t_{\min} = \SI{1E-9}{\second}$, respectively, and a viscosity $\eta_\nv{f} = \SI{e-5}{\kilo\newton\per\milli\meter\squared\per\second}$ is applied. In order to investigate the impact of this numerically motivated parameter, a convergence study has been carried out. For this purpose, $\eta_\nv{f} =\SI{e-6}{\kilo\newton\per\milli\meter\squared\per\second}$ and $\eta_\nv{f} = \SI{e-4}{\kilo\newton\per\milli\meter\squared\per\second}$ were considered. The viscosity did not have an influence on the obtained crack pattern, nor did it affect the results which were taken as quantification like the energy release rate or crack tip position.

\subsection{Homogeneous elasticity and crack perpendicular to interface}
\label{sec:elhomperp}

In a first numerical study, a large variety of different fracture toughness ratios $\gcb/\gci \in (1,10]$ has been considered.
For all functions $\rheavi$, $\rgauss$ or $\rexpo$, and different values of the interface width $\elli$, the crack patterns have been simulated and investigated.
A constant \textsc{Young}'s modulus $E=\SI{210}{\kilo\newton\per\milli\meter\squared}$ and perpendicular interface, $\varphi_\text{i}=\SI{90}{\degree}$, are considered to keep the setup as simple as possible.
Table~\ref{tab:ReginvRes} gives a representative selection of the interface descriptions and fracture toughness ratios which have been investigated and presents the corresponding simulation results.

\begin{figure}[t]
	\centering
	\subfloat[$\gcb/\gci=4.5$]{
		\includegraphics[width=0.30\textwidth]{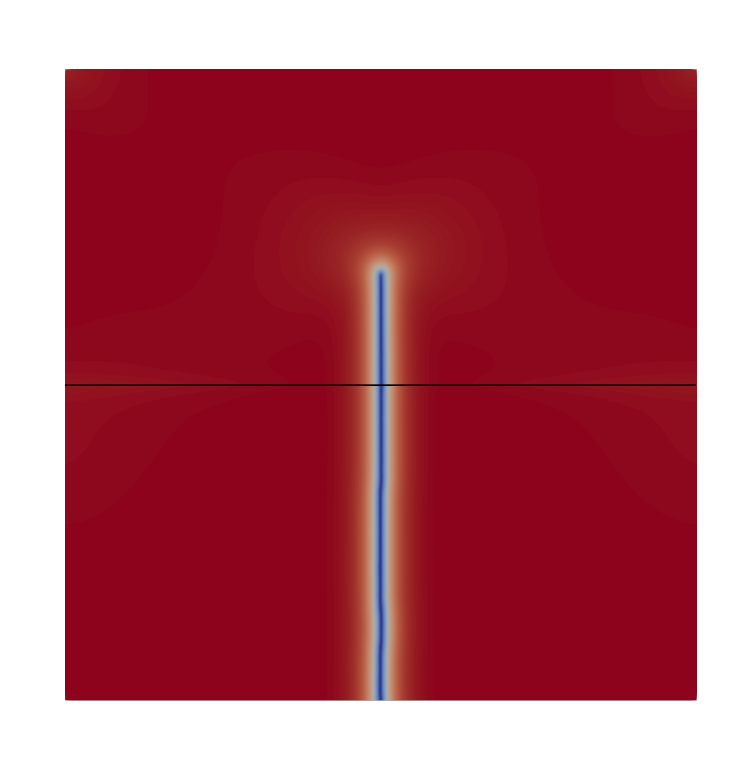}
		\label{fig:contPhen_str}
	}
	\subfloat[$\gcb/\gci=4.75$]{
		\includegraphics[width=0.30\textwidth]{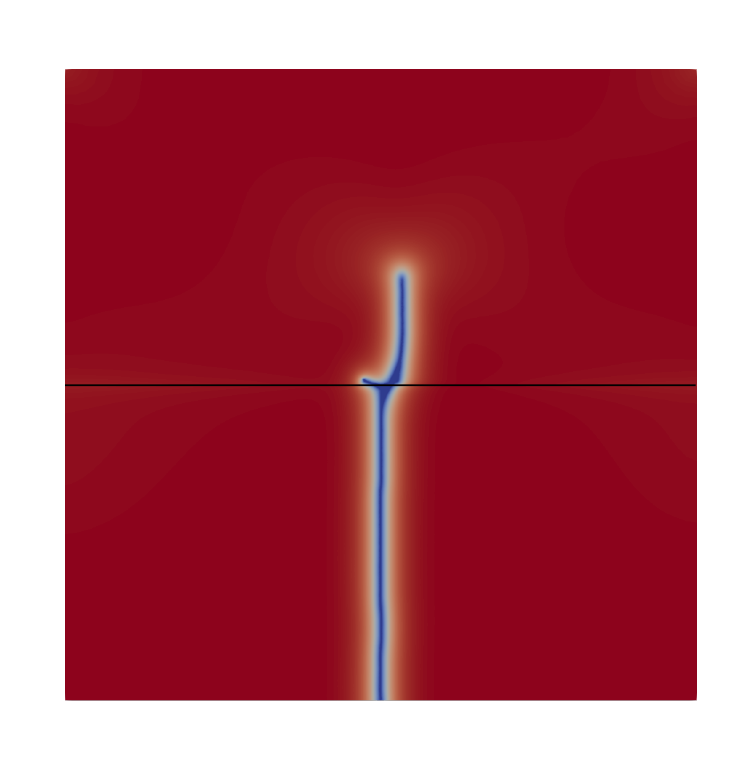}
		\label{fig:contPhen_ki}
	}
	\subfloat[ $\gcb/\gci=6$]{
		\includegraphics[width=0.30\textwidth]{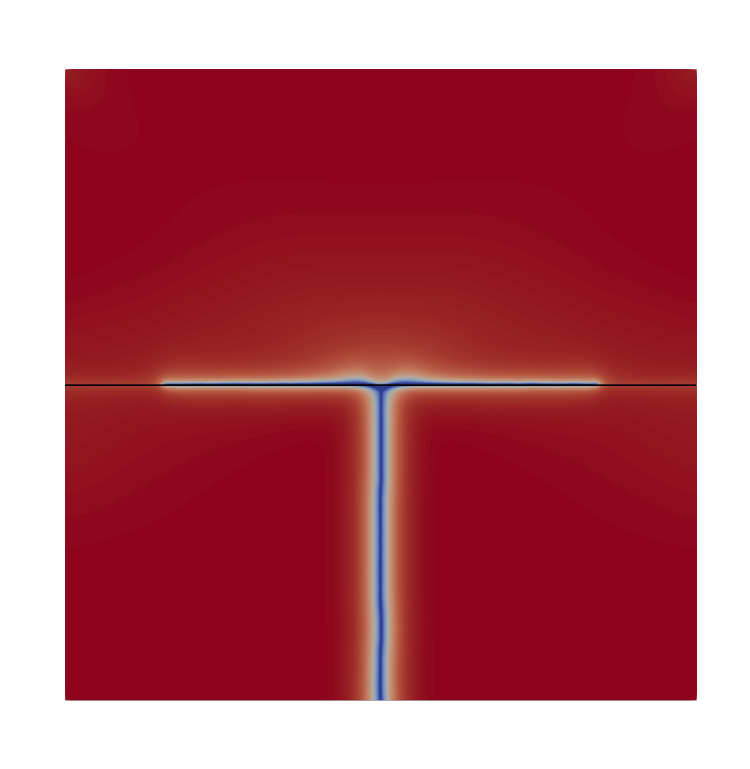}
		\label{fig:contPhen_br}
	}
	\raisebox{0.8cm}{\includegraphics{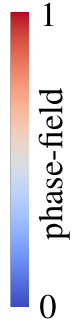}}
	\caption{Three different crack phenomena for an initial crack perpendicular to the interface which is depicted using $\rgauss$ with $\elli/\ellc=1.25$. The size of the contour plots corresponds to the domain size.
	Similar crack phenomena are obtained for other interface descriptions, i.e. for $\rheavi$ and $\rexpo$ as well as different interface widths $\elli$.
	The interface midline is indicated in black. The phenomena \textbf{(a)}~--~\textbf{(c)} correspond to the last three columns in Table~\ref{tab:ReginvRes}.}
	\label{fig:contPhen}
\end{figure}

The numerically predicted crack patterns can be divided into three groups. Representative examples for these three phenomena are depicted in Figure~\ref{fig:contPhen}:

\begin{enumerate}[label=(\alph*)]

\item For fracture toughness ratios lower than a critical value $\gcb/\gci$, crack growth straight across the interface is induced, Figure~\ref{fig:contPhen_str}.
In other words, there is no influence of the interface regarding the crack path.

\item Considering higher ratios of the fracture toughnesses $\gcb/\gci$, crack branching occurs when the crack approaches the interface midline and a symmetric growth with respect to the $y$-axis is observed.
For some interface descriptions and ratios $\gcb/\gci$, one of the two crack tips kinks into the bulk material beyond the interface when the interfacial crack advanced a bit, see Figure~\ref{fig:contPhen_ki}.
The choice whether it is the left or right tip is governed by numerical round-off errors. As soon as one of the crack tips kinks into the adjacent bulk material, the other crack tip arrests and does not propagate for the rest of the simulation.

\item For higher $\gcb/\gci$ and some interface descriptions, the crack branches into the interface, yet no subsequent kinking into the bulk material appears. Accordingly, a crack arises along the interface, approaching the vertical edges of the domain for a large simulation time, Figure~\ref{fig:contPhen_br}.

\end{enumerate}

\begin{table}[tb]
    \centering
    \caption{Crack phenomena at an interface perpendicular to the initial crack for different regularization functions and length scales. The \textsc{Young}'s modulus is constant within the entire domain. The results are a representative selection of those obtained for various fracture toughness ratios $\gcb/\gci \in (1,10]$.
    According to LEFM \cite{he_crack_1989}, crack growth straight across the interface is expected for $\gcb/\gci \lesssim 4$, branching into the interface for $5 \lesssim \gcb/\gci$. 
    For $4 \lesssim \gcb/\gci \lesssim 5$, a single deflection into the interface is analytically predicted which was not recovered in any simulation.}
    \label{tab:ReginvRes}
    \begin{tabular}{c|c||c|c||c|c|c}
   \multicolumn{2}{p{2cm}||}{Interface \mbox{description}} & $\gcb/\gci$ & $\gcb/\hat{\mathcal{G}}_\text{c}^\text{i}$ & \multicolumn{1}{p{1.9cm}|}{\multirow{2}{2.1cm}[0mm]{crack growth straight across the interface}} &\multicolumn{2}{c}{branching into the interface \mbox{followed} by \dots} \\ 
   $\gc$ & $\elli/\ellc$&&&& \multicolumn{1}{p{2.8cm}|}{\dots kinking into bulk}& \multicolumn{1}{p{2.9cm}}{\dots interfacial failure} \\[2mm] 
        \hline 
	  \multicolumn{1}{c|}{\multirow{8}[0]{*}{$\rheavi$}} &
	  \multicolumn{1}{c||}{\multirow{3}[0]{*}{$2.5$}} &
		  4.2    &   4.62   &  x      &       & \\
		         & & 4.8  &    5.33  &       & x &  \\
      &  & 6  &    6.76  &       & x &  \\
       & & 8  &    9.18  &       &  & x \\[2mm]

	&  \multicolumn{1}{c||}{\multirow{2}[0]{*}{$1.25$}} &
                    3.3  &   4.78    &   x    &       &  \\
		& &    4.2   &   6.56    &       &    x    &   \\
				& &    5.1  &  8.47    &       &        & x  \\[2mm] 
    \hline 
	  \multicolumn{1}{c|}{\multirow{12}[0]{*}{$\rgauss$}} &
	  \multicolumn{1}{c||}{\multirow{3}[0]{*}{$2.5$}} &
		  5     &    8.38  &  x      &       & \\              
                & & 7   &   9.63 &   x    &  &  \\
                & & 8  &    11.58  &       &  & x \\[2mm]
         & \multicolumn{1}{c||}{\multirow{4}[0]{*}{1.875}} & 5   &   7.12 &    x   &       &  \\
              &  &  6     &    9.34   &  x      &       & \\
	&	&  6.5     &    10.55   &       &       &x \\
                & &  7     &    11.82   &       &       & x \\[2mm]

	&  \multicolumn{1}{c||}{\multirow{4}[0]{*}{$1.25$}} &
                    4.5   &   8.38    &   x    &       &  \\
		& &    4.75   &  9.26    &        &   x   &  \\
         &       &    5.5   &   12.18    &       &     x   &  \\
		& &    6   &   14.31    &       &        &  x \\[2mm] 
	\hline 
	  \multicolumn{1}{c|}{\multirow{6}[0]{*}{$\rexpo$}} &
	  \multicolumn{1}{c||}{\multirow{3}[2]{*}{$3.125$}} &
		  5.5     &    12.25  &  x      &       & \\              
       & & 6  &    15.46  &   x    &   &  \\
              & & 6.5  &    19.88  &       &  x &  \\[2mm]
              
              	  & \multicolumn{1}{c||}{\multirow{3}[2]{*}{$2.5$}} &
		  5     &    12.66  &  x      &       & \\              
       & & 5.5  &    17.19  &       &  x &  \\
              & & 6  &    24.6  &       &  x &  \\[2mm]
    \end{tabular}
\end{table}

From Table~\ref{tab:ReginvRes}, it becomes clear that the description of the interface, i.e. the choice of the function $\rheavi$, $\rgauss$ or $\rexpo$, and the value of the interface width $\elli$ can have an impact on the simulation result although the procedure outlined in Section~\ref{sec:compappr} generally allows for energetically equivalent crack propagation along differently regularized interfaces.

Apparently, for a constant $\elli$, the choice of the regularization function can affect the numerically predicted crack path. The critical ratio $\gcb/\gci$ for crack branching into the interface increases from the \textsc{Heaviside}-like to the exponential to the \textsc{Gaussian}-like regularization, cf. rows with $\elli/\ellc=2.5$ and column $\gcb/\gci$.
In contrast, the compensated fracture toughness ratio $\gcb/\gcih$ increases from the \textsc{Heaviside}-like to the \textsc{Gaussian}-like to the exponential regularization. Thus, the impact of the choice of the regularization function on the results seems not only to be caused by the compensation procedure but also by the regularization directly.

A variation of the interface width $\elli$ influences the crack path for all the regularization functions in a similar way: In the context of crack deflection, an interface of higher $\elli$ seems to be tougher than a narrower one.
For example, for $\rgauss$ crack branching into the interface occurs for $4.5 \lesssim \gcb/\gci$ when $\elli/\ellc=1.25$, while $8 \lesssim \gcb/\gci$ has to be reached if $\elli/\ellc=2.5$ is set.
Furthermore, it depends on the interface length scale and the regularization function, respectively, whether the crack propagates within the interface or kinks out into the bulk material when it has branched into the interface.
Arguably, these effects are triggered by different $\elli$ and not by the compensation. If only the compensation procedure would have an influence, one would expect monotonously rising ratios $\gcb/\gcih$ for the transition between the phenomena from higher to lower $\elli$ values because the bulk material influence rises. This is, however, not the case, as can be seen from Table~\ref{tab:ReginvRes} when comparing the compensated ratios for $\gcb/\gcih$ corresponding to the critical ratios for crack deflection $\gcb/\gci$ for the \textsc{Gaussian} regularization for different length scale ratios.

Due to its impact on the crack path, the interface width $\elli$ may not be regarded as a purely numerical parameter.
Rather, it should be considered as a material parameter in addition to the fracture toughness $\gci$. In other words, $\elli$ can be assigned an experimentally determined value.
This is consistent with experimental investigations of Park and Chen~\cite{park_experimental_2011}, and Parab and Chen~\cite{parab_crack_2014}. 
In both papers, dynamic crack propagation is investigated within two brittle solids linked by an interface. The interface has a varying, finite width and is composed of an adhesive. Depending on the interface width, different crack patterns can arise.
In this paper, numerical results obtained with the regularized model have been compared to LEFM investigations, which assume an infinitesimal narrow interface.
Hence, a physically motivated choice of the regularization width $\elli$ is beyond the scope of this work. Instead, $\elli$ is set such that optimal agreement is obtained between the simulation and the analytical results.

According to LEFM investigations \cite{he_crack_1989}, straight crack growth across the interface is expected for \mbox{$\gcb/\gci \lesssim 4$}. For interfaces of low fracture toughness, $5 \lesssim \gcb/\gci$, crack branching into the interface is energetically favorable.
Hence, the numerically predicted crack phenomena are in good qualitative agreement with LEFM predictions.
However, He and Hutchinson~\cite{he_crack_1989} anticipated a single deflection effect, i.e. crack deflection into the interface not coming along with branching, for fracture toughness ratios $4 \lesssim \gcb/\gci \lesssim 5$.
This crack path has not been found in the simulations. Instead, a different asymmetric phenomenon, crack branching followed by a kink into the bulk material, has been predicted.
This limitation may be due to the application of the tensile split \eqref{eq:tenssplit} which is not capable of fully degrading materials under non-mode-I loading.
A remedy may be the directional split~\cite{steinke_phase-field_2018} which degrades the individual components of the stress tensor on a physical basis.

In terms of convergence, the \textsc{Gaussian}-like and the exponential regularization appear to have an advantage, which is why the \textsc{Heaviside}-like description is not considered in the following investigations.
For the exponential regularization $\rexpo$, a pronounced influence of the bulk material on the interface fracture toughness was observed in the study outlined in Section~\ref{sec:compappr}.
This leads to a strong saturation effect. In other words, the fracture toughness ratio $\gcb/\gci$ that can be reached is limited to a rather small value.
For example, a maximum value of $\gcb/\gci \approx 2.7$ can be estimated from Figure~\ref{fig:compensation_e} for $\elli/\ellc=0.625$: Even a ratio of $\gcb/\gcih=50$, which is not shown in the figure, yielded $\gcb/\gci < 2.7$.
This leads to strong limitations concerning the crack phenomena which can be captured.
Accordingly, the \textsc{Gaussian}-like regularization is used in the remainder of the paper. It is applied with $\elli/\ellc=1.25$, as an optimal accordance between the regularized interface model and the results from LEFM is obtained in this way.

\begin{figure}[t]
	\centering
	\subfloat[Energy release rate, $\gcb/\gci=3$]{
		\includegraphics{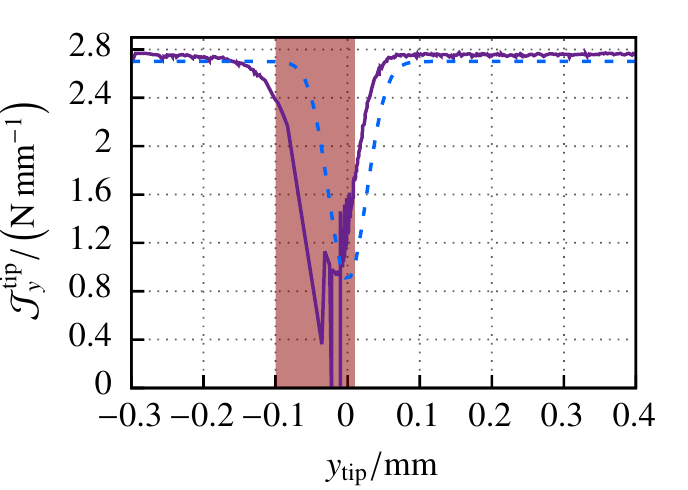}
		\label{fig:quantAn_str_err}
	}
	\subfloat[Crack tip trajectory, $\gcb/\gci=3$]{
		\includegraphics{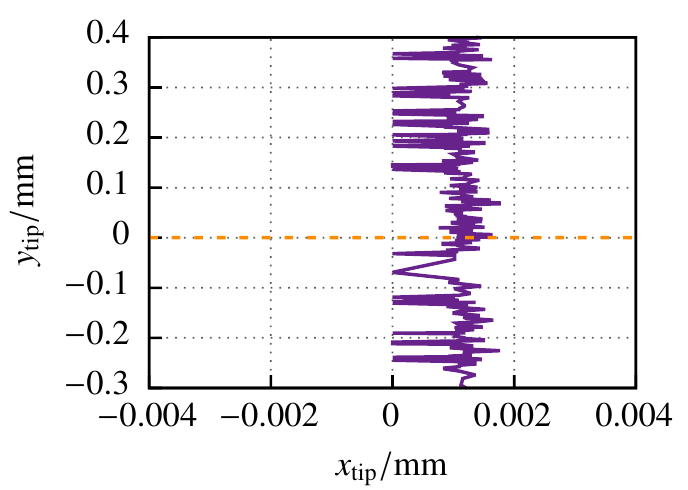}
		\label{fig:quantAn_str_tip}
	}
	
	\subfloat[Energy release rate, $\gcb/\gci=8$]{
		\includegraphics{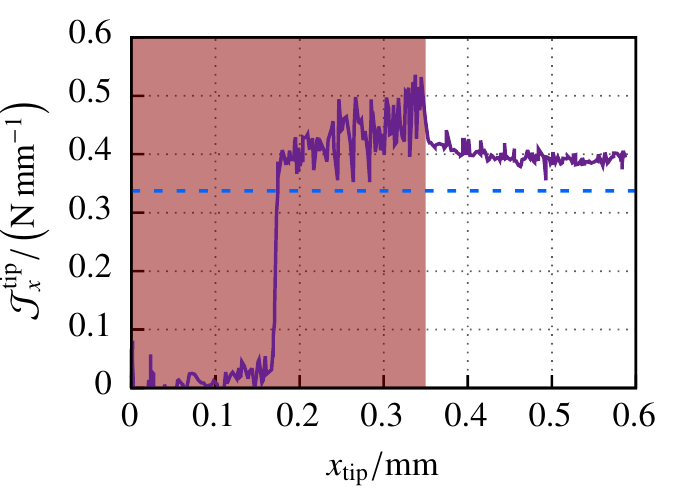}
		\label{fig:quantAn_br_err}
	}
	\subfloat[Crack tip trajectory, $\gcb/\gci=8$]{
		\includegraphics{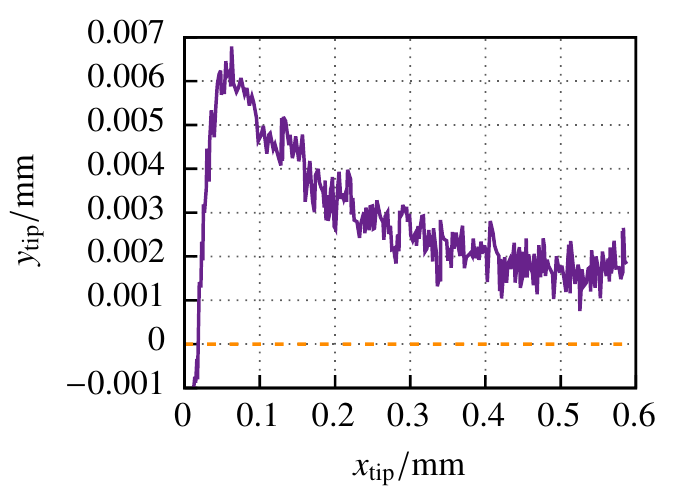}
		\label{fig:quantAn_br_tip}
	}
	\caption{Energy release rates (left) and crack tip trajectories (right) for a straight (top,  cf. Figure~\ref{fig:contPhen_str}) and a branched (bottom, cf. Figure~\ref{fig:contPhen_br}) crack. The energy release rates for a straight \textbf{(a)} and branched \textbf{(c)} crack are compared to the fracture toughness $\rgauss$ (\legline{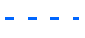}) along the $y$- and $x$-axis, respectively. The red intervals correspond to oscillations in \textbf{(a)} and no valid evaluation in \textbf{(c)}. Outside these intervals, there is good agreement between the energy release rate, at which the crack propagates and the fracture toughness.
	The corresponding crack tip trajectories are shown in \textbf{(b)} and \textbf{(d)}, where \legline{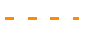} marks the interface midline. The trajectories are, as expected, either straight along the $y$-axis \textbf{(b)} or picture deflection into the interface along the $x$-axis \textbf{(d)}. In both cases, small deviations occur due to the crack tip tracking method. For the deflected crack, there is still a tendency to penetrate into the adjacent bulk material layer, which is why the crack is not exactly centered in \textbf{(d)}.}
	\label{fig:quantAn}
\end{figure}

So far, the ability of the model to predict failure phenomena which are consistent with LEFM has been demonstrated.
A deeper insight into the effect of the regularization on the crack driving forces is obtained by consulting the energy release rate which is determined from the balance of the configurational forces and the crack tip trajectory.
The corresponding curves for a crack growing straight across the interface or propagating along the interface, respectively, are presented in Figure~\ref{fig:quantAn}.
As cracks are described in a regularized manner, the definition of a discrete crack tip is not a trivial question.
Here, all nodes with a phase-field value $c<c_\nv{th}$, i.e. lower than the critical threshold $c_\nv{th}$ introduced in the context of the irreversibility constraint in Section~\ref{sec:diffeqcdu}, are considered. Then, the furthest top right node is identified as the actual tip.

Figure~\ref{fig:quantAn_str_tip} depicts the crack tip trajectory for a fracture toughness ratio $\gcb/\gci=3$.
The crack grows straight across the interface, i.e. propagates symmetrically along the $y$-axis. Its $x$-coordinate $x_\nv{tip}$ is slightly overestimated, because of the crack tip tracking method explained in the previous paragraph.
The corresponding energy release rate $\jt_y$ is shown in Figure~\ref{fig:quantAn_str_err}. Away from the interface, a value equal to the fracture toughness of the bulk material $\gcb$ is recovered, which is expected.
Closer to the interface midline, $\jt_y$ follows the regularization function $\rgauss$. However, significant deviations occur when the crack tip approaches the interface, i.e. for \mbox{$\SI{-0.1}{\milli\meter} \lesssim y_\nv{tip} \lesssim \SI{0.01}{\milli\meter}$}. The corresponding interval is indicated in red in Figure~\ref{fig:quantAn_str_err}.
Comparable deviations of the energy release rate or oscillations, respectively, are observed in all simulations. As these do not coincide for two simulations with an identical setup, they are assumed to be caused by numerical errors arising from the evaluation of the configurational force balance in FEniCS. Such obscure phenomena never occurred in the alternative finite element environment of the authors, were not reported in literature using the same approach~\cite{kuhn_discussion_2016} and question the usability of FEniCS for complex constitutive models. 

For $\gcb/\gci=8$, the crack branches into the interface when approaching its midline. The trajectory of the right crack tip propagating in the positive $x$-direction is depicted in Figure~\ref{fig:quantAn_br_tip}. The crack tip overshoots the interface midline at the beginning, but follows a curved path and approaches the midline when it continues to propagate in $x$-direction. The elastic energy, that has to be built up to propagate the crack towards the interface through the bulk material with a higher fracture toughness is suddenly released. The crack snaps into the interface and the elastic energy, which is released, suffices for the crack to tackle the first bit of the energetic barrier towards the second bulk material layer. However, as the simulation continues, it is energetically more favourable for the crack to find its path closer to the interface midline, where the deviation at $x_\text{tip}=\SI{0.5}{\milli\meter}$ is almost the same as for the straight crack.
In contrast to a sharp interface model, the crack propagating along the regularized interface does not follow the interface midline exactly.
Nevertheless, the uncertainty arising from the regularization and the tracking method does not exceed $\elli/2$, which is deemed acceptable regarding the characteristic domain size $a$. 
Figure~\ref{fig:quantAn_br_err} shows the corresponding energy release rate $\jt_x$. It is noted that the validity of $\jt_x$ determined from the configurational force balance is compromised for $x_\nv{tip} < r = \SI{0.35}{\milli\meter}$. This is due to the crack tip propagating in the opposite direction along the interface and the point of crack branching which are located within the integration domain $\mathcal{C}$ of the configurational forces. The former cancels out $\jt_x$ for $0 \leq x_\nv{tip} < r/2$ and the latter provokes oscillations within $r/2 \leq x_\nv{tip} \leq r $.
The corresponding interval is marked in red in Figure~\ref{fig:quantAn_br_err}.
Only for $x_\nv{tip} > r = \SI{0.35}{\milli\meter}$, $\jt_x$ recovers the actual crack driving force.
The energy release rate of the crack propagating along the interface approximately meets the interface fracture toughness $\gci$, yet is slightly higher than the exact value.
On the one hand, this slight overestimation of $\gci$ is due to the crack tip not propagating exactly along the interface midline, an issue due to the interface regularization. This has already been outlined in Section~\ref{sec:compappr} as a limitation.
On the other hand, $\jt_x$ exhibits an uncertainty which stems from the definition of a discrete crack tip position based on the phase-field. 

It is noted that a higher ratio $\elli/\ellc$ can lead to a larger deviation of the crack tip from the interface midline. This may result in a larger discrepancy between the actual driving force of a crack propagating along the interface and the interface fracture toughness.
Hence, the interface length scale $\elli$ should not be chosen significantly larger than the crack regularization length $\ellc$.

\subsection{Heterogeneous elasticity and crack perpendicular to interface}
\label{sec:hetel-per}
\begin{figure}[t]
	\centering
	\subfloat[Elastic heterogeneity regularized using $\rtan$]{
		\includegraphics{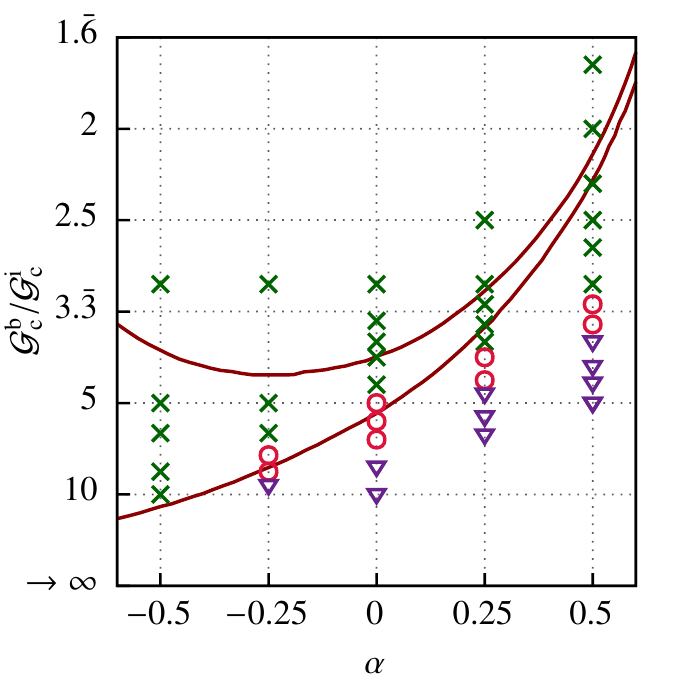}
		\label{fig:modelResRect_reg}
	}
	\subfloat[Conforming jump of the \textsc{Young}'s modulus]{
		\includegraphics{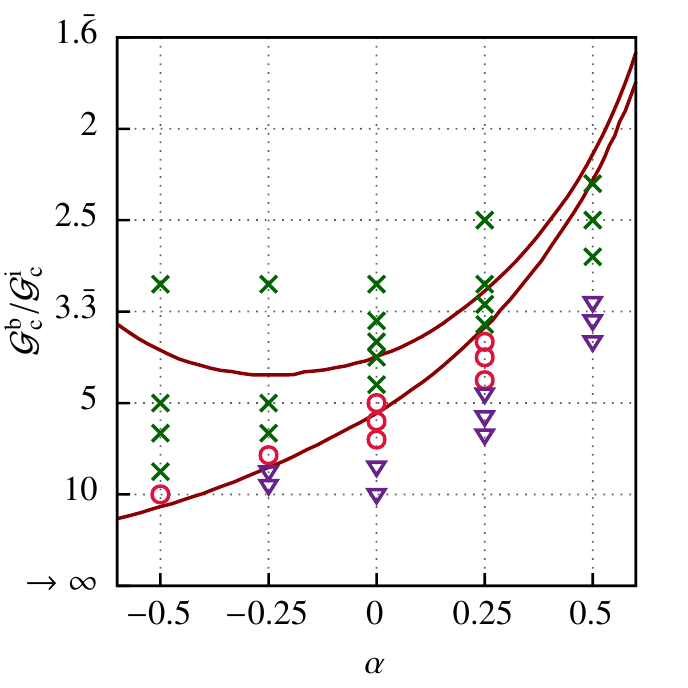}
		\label{fig:modelResRect_jmp}
	}
	\\
	\vspace{0.3cm}
	\footnotesize 
	\begin{tabular}{l|l}
		\legpoint{1} Straight crack across interface & \legpoint{5} Branching followed by kink into adjacent bulk material\\ 
		\legpoint{9} Branching and propagation along interface & \legline{3} Analytically predicted switch between phenomena \\   
	\end{tabular}
	\caption{Numerical results for a regularized \textbf{(a)} and a conforming jump \textbf{(b)} of the \textsc{Young}'s modulus are compared to predictions made by LEFM~\cite[Figure 3]{he_crack_1989}. \textbf{Above the red lines}, LEFM predicts straight crack growth across the interface. \textbf{In between}, a single deflection is expected. \textbf{Below the red lines}, a double deflection should appear. The numerical results are in good agreement with the analytical predictions. However, instead of a single deflection the crack keeps growing straight across the interface. It is noted that the \textit{branching} and the \textit{branching followed by kinking} phenomena, denoted by a triangle and a circle respectively, count as a double deflection in the context of LEFM.
	When comparing \textbf{(a)} to \textbf{(b)}, it can be seen, that the regularization of the elastic dissimilarity influences the results.}
	\label{fig:modelResRect}
\end{figure}

Elastic heterogeneities can have a crucial influence on the failure phenomena which arise when a crack approaches an interface. The dissimilarity of the elastic fields on each side of the interface can induce crack patterns that differ from those which occur in the case of homogeneous elastic constants.
He and Hutchinson \cite{he_crack_1989} argued for a setup similar to the one investigated here, that the consequences of the elastic heterogeneity can be characterized by a dimensionless parameter $\alpha$ introduced by Dundurs~\cite{dundurs_discussion:_1969}. For the plane strain setting and a constant \textsc{Poisson} ratio $\nu$ which are assumed in this paper, the first \textsc{Dundurs}' parameter~$\alpha$ may be written as a function of the \textsc{Young}'s moduli of the bulk material,
\begin{equation}
    \alpha = \frac{E_2-E_1}{E_1+E_2} \coma
    \label{eq:alpha}
\end{equation}
in which $E_1$ and $E_2$ refer to the bulk in front of and beyond the interface, respectively.
In order to analyze the effect of the elastic dissimilarity on crack propagation, another numerical study has been carried out. Therefore, the \textsc{Young}'s modulus $E_2$, assigned to the material beyond the interface, has been varied and four different\linebreak values~\mbox{$\alpha \in \{-0.5, -0.25, 0.25, 0.5 \}$} were investigated.
In the first part of the study, the elastic heterogeneity is captured by the hyperbolic tangent function $\rtan$. In order to investigate if $\rtan$ has a significant impact on the predicted crack path, a conforming jump $\reheavi$ of the \textsc{Young}'s modulus with respect to the interface midline is considered, subsequently.
The \textsc{Gaussian}-like regularization $\rgauss$ with $\elli/\ellc=1.25$ has been applied. All other parameters are identical to the values in the previous section.
Various fracture toughness ratios $\gcb/\gci \in \left[1.5,10\right]$ are considered.

A representative selection of the parameters investigated and the corresponding results are depicted in Figure~\ref{fig:modelResRect}. Therein, relevant results for the case of a homogeneous \textsc{Young}'s modulus, i.e. $\alpha=0$, are duplicated from Section~\ref{sec:elhomperp}.
The result of every simulation matches one of the crack phenomena described in the previous section. The crack grows either straight across the interface or it branches into the interface. For some branching cases, one of the two crack tips kinks out of the interface when the interfacial crack advanced a bit, while the other one is arrested.

The critical fracture toughness ratio which has to be reached for crack branching into the interface decreases with increasing values of $\alpha$. Thus, crack propagation within the interface becomes energetically more favourable when the material beyond the interface is stiffer.
This is consistent with the LEFM predictions \cite{he_crack_1989}. Furthermore, the results approximate the analytically predicted correlation between this critical fracture toughness ratio and the parameter $\alpha$ especially for $\alpha \leq 0.25$. However, a significant deviation appears for $\alpha=0.5$. 

Comparing the crack patterns with a regularized jump $\rtan$ to those in which a mesh-conforming jump of the \textsc{Young}'s modulus $\reheavi$ has been considered, see Figures~\ref{fig:modelResRect_reg} and \ref{fig:modelResRect_jmp}, it becomes obvious that the incorporation of the elastic dissimilarity significantly influences the numerically predicted crack pattern.
This may be due to the fact that the use of $\rtan$ does not necessarily lead to a solution which satisfies the mechanical jump conditions, so that unphysical values of the strain energy can occur in the vicinity of the interface, cf.~\cite{schneider_phase-field_2015,kiefer_numerical_2017}.
This is a clear limitation of the model presented herein.
A remedy to this issue is for example the \textit{partial rank-I relaxation}~\cite{mosler_novel_2014,schneider_phase-field_2015,kiefer_numerical_2017}.

\subsection{Homogeneous elasticity and inclined interface}
In order to generalize the previous findings for an interface perpendicular to an initial crack, the setup is extended to an inclined interface. As an example, the study of three inclination angles \mbox{$\phii \in \lbrace\SI{30}{\degree},\SI{45}{\degree}, \SI{60}{\degree}\rbrace$} and various fracture toughness ratios $\gcb/\gci \in (1,5]$ is presented.
Within this section, homogeneous elastic constants are considered. All other parameters are as in the previous section.

He and Hutchinson \cite{he_crack_1989} considered a crack which impinges an inclined interface and analytically determined the ratio
\begin{equation}
     \frac{\gb}{\gi} = 16 \, \left( \left[ 3 \, \cos\left(\frac{\phii}{2} \right) + \cos\left(3 \, \frac{\phii}{2} \right)  \right]^2
 + \left[ \sin\left(\frac{\phii}{2} \right) + \sin\left(3 \,  \frac{\phii}{2} \right)  \right]^2 \right)^{-1} 
    \label{eq:err-an-incl}
\end{equation}
of the energy release rates for straight crack propagation across the interface~$\gb$ and crack deflection into the interface~$\gi$, cf. \cite[Equation (36)]{paggi_revisiting_2017}.
Deflection into the interface is expected to occur for $\gcb/\gci>\gb/\gi$. In other words, the ratio~$\gb/\gi$ is identical to what has been named the critical fracture toughness ratio in the previous sections.
It is remarked that, in contrast to the analytical investigations which serve as comparison, a finite distance between the initial crack $\gammalc_0$ and the interface is considered for the simulations, see Figure~\ref{fig:simsetup}. Although this is a difference to the analytical reference, a crack which has to approach an interface from a finite distance first is considered for the numerical investigations, since this is the more realistic and more general case, while crack nucleation within the interface is not discussed in this contribution. 

\begin{figure}[t]
	\centering
	\subfloat[$\gcb/\gci=1.6$]{
		\includegraphics{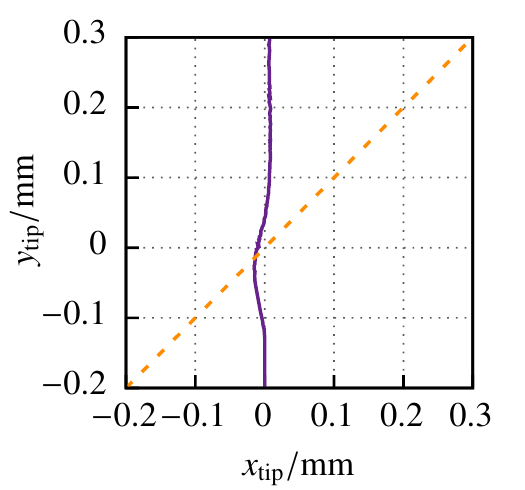}
		\label{fig:contIncl_defl1}
	}
	\subfloat[$\gcb/\gci=2.4$]{
		\includegraphics{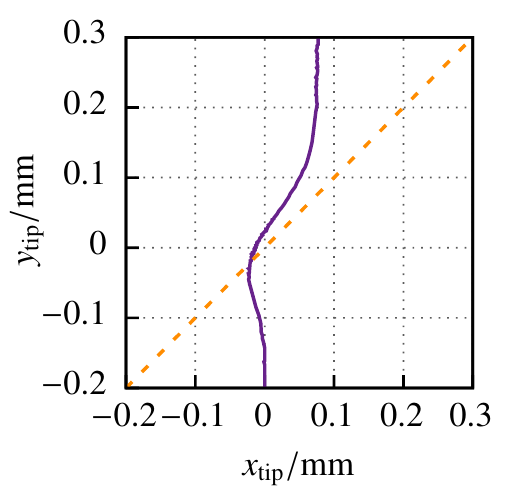}
		\label{fig:contIncl_defl2}
	}
	\subfloat[$\gcb/\gci=3$]{
		\includegraphics{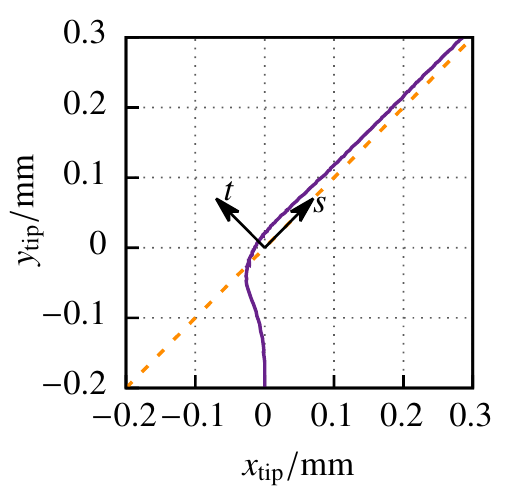}
		\label{fig:contIncl_defl3}
	}
	\caption{Three different crack phenomena for an initial crack which propagates towards an inclined interface $\phii=\SI{45}{\degree}$, depicted by the midline~\legline{8}. Depending on the ratio $\gcb/\gci$, the crack tip trajectory experiences almost no \textbf{(a)} to significant \textbf{(c)} influence of the interface. Unlike the sharp transition between a straight and deflected crack for the critical ratio assumed in \cite{he_crack_1989}, the crack patterns exhibit a transition from a straight to a deflected crack.}
	\label{fig:contIncl}
\end{figure}

For all inclination angles $\phii$, similar crack phenomena are predicted numerically, depending on the fracture toughness ratio $\gcb/\gci$.
For $\phii=\SI{45}{\degree}$ and three representative fracture toughness ratios, the crack tip trajectories obtained from simulations are depicted in Figure~\ref{fig:contIncl}.
In general, the crack does not follow a straight path when approaching the interface. Instead, it is deflected along a curved path towards the interface.
This deflection is more pronounced for higher values of the fracture toughness ratio~$\gcb/\gci$ and smaller values of the interface inclination~$\phii$.
It is noted that this deflection leads to a discrepancy of the actual angle between interface and crack, and the initial inclination angle $\phii$.
Thus, the significance of the prediction \eqref{eq:err-an-incl} when using the initial inclination angle can be biased for the numerical setup considered here.

For lower values of the fracture toughness ratio $\gcb/\gci$ the crack propagates across the interface into the second material, see Figure~\ref{fig:contIncl_defl1}. Within the second material layer, it firstly follows a curved path again, yet in the opposite $x$-direction with respect to the path it took when approaching the interface. The path is shaped such that the crack continues to propagate approximately vertically when it reaches the $y$-axis, i.e. it further propagates aligned with the initial crack.

\begin{figure}[t]
	\centering
	\subfloat[$\phii=\SI{30}{\degree}$]{
		\includegraphics{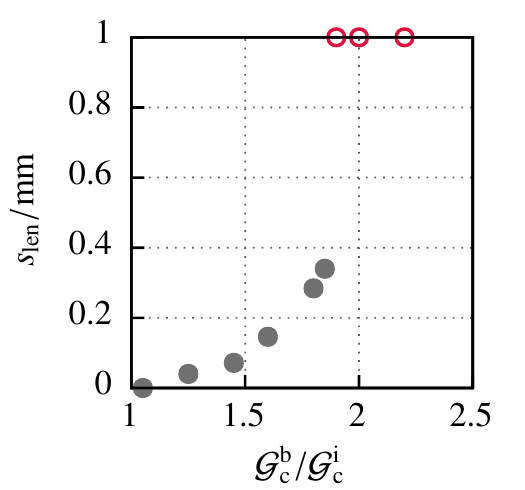}
		\label{fig:crackLenIF_30}
	}
	\subfloat[$\phii=\SI{45}{\degree}$]{
		\includegraphics{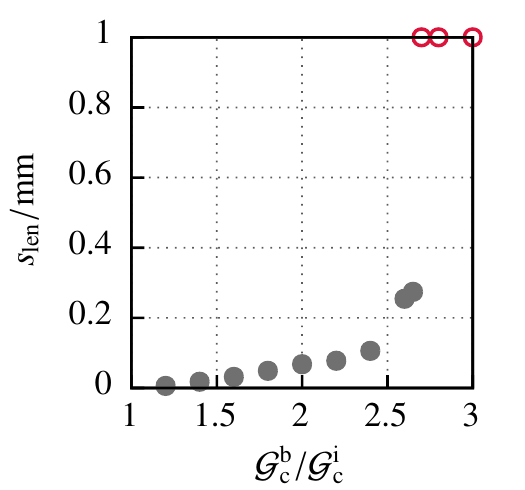}
		\label{fig:crackLenIF_45}
	}
	\subfloat[$\phii=\SI{60}{\degree}$]{
		\includegraphics{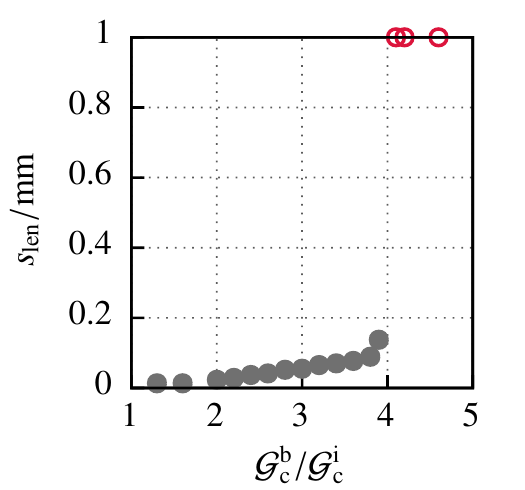}
		\label{fig:crackLenIF_60}
	}
	\\
	\vspace{0.3cm}
	\footnotesize 
	\begin{tabular}{l|l}
		\legpoint{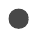} $s_\text{len}<\SI{1}{\milli\meter}$ & \legpoint{5} $s_\text{len}\geq\SI{1}{\milli\meter}$\\
	\end{tabular}
	\caption{Crack length along the interface $s_\text{len}$ for three different inclination angles \textbf{(a)}~--~\textbf{(c)} and various ratios $\gcb/\gci$. As expected, $s_\text{len}$ recovers higher values for larger angles $\phii$ and constant ratios $\gcb/\gci$. Additionally, the ratio, where the crack deflects into the interface gets more and more pronounced with rising $\phii$. In general, a rather smooth transition is observed for smaller inclination angles, which makes a comparison to the analytic results, where a sharp transition is predicted, difficult.}
	\label{fig:crackLenIF}
\end{figure}

For higher values of the fracture toughness ratio $\gcb/\gci$, the crack deflects into the interface. However, there is no sharp transition between interfacial failure and crack penetration into the bulk material beyond the interface.
Instead, for intermediate ratios $\gcb/\gci$, the crack propagates along the interface for a length $s_\text{len}$\footnote{For the definition of $s_\text{len}$, the crack tip coordinates $[s_\text{tip}\,\, t_\text{tip}]\T$ are evaluated in the $s,t$-coordinate frame aligned with the interface, cf. Figure~\ref{fig:contIncl_defl3}. In the context of the diffuse interface representation, the crack is assumed to propagate along the interface for $|t| < \elli$.}
which is higher for weaker interfaces with respect to the bulk, and for smaller angles $\phii$, cf. Figure~\ref{fig:crackLenIF}. Subsequently, it kinks out into the material beyond the interface where it continuous to grow parallel to the $y$-axis, see Figure~\ref{fig:contIncl_defl2}.
When a certain value of the ratio $\gcb/\gci$ is reached, no more kinking out of the interface has been observed, see Figure~\ref{fig:contIncl_defl3}.
Both, the increase of $s_\text{len}$ with increasing $\gcb/\gci$ and decreasing $\phii$, and the increase of $\gcb/\gci$, for which no more kinking out of the interface occurs, for increasing $\phii$ are consistent with the numerical results from \cite{paggi_revisiting_2017} and the analytical reference \cite{he_crack_1989} which predicts the increase of $\gb/\gi$ for increasing $\phii$, cf. Equation~\eqref{eq:err-an-incl}.

Since the results indicate rather a smooth transition between failure of the bulk material beyond the interface and interfacial rupture than a sudden switch between the two phenomena, the determination of a critical fracture toughness ratio from the simulation results and its comparison to LEFM predictions, respectively, are not simple. However, considering Figures~\ref{fig:crackLenIF_45} and \ref{fig:crackLenIF_60}, $s_\text{len}$ remarkably increases for $\phii=\SI{45}{\degree}$, when $\gcb/\gci \gtrsim 2.4$ and for $\phii=\SI{60}{\degree}$, when $\gcb/\gci \gtrsim 3.8$. In contrast, the analytically predicted ratios are $\gb/\gi = 1.37$ and $\gb/\gi = 1.78$, respectively. This discrepancy suggests, that the prescribed angle $\phii$ is not decisive for interface failure, but the angle between the interface and the crack tip trajectory, when the crack has already turned towards the interface. Considering Figure~\ref{fig:contIncl_defl2}, this angle is approximately $\SI{60}{\degree}$. Using this value, the ratio $\gb/\gi = 1.78$ still underestimates $\gcb/\gci \gtrsim 2.4$. 

\begin{figure}[t]
	\centering
	\subfloat[Energy release rate, $\gcb/\gci=3$]{
		\includegraphics{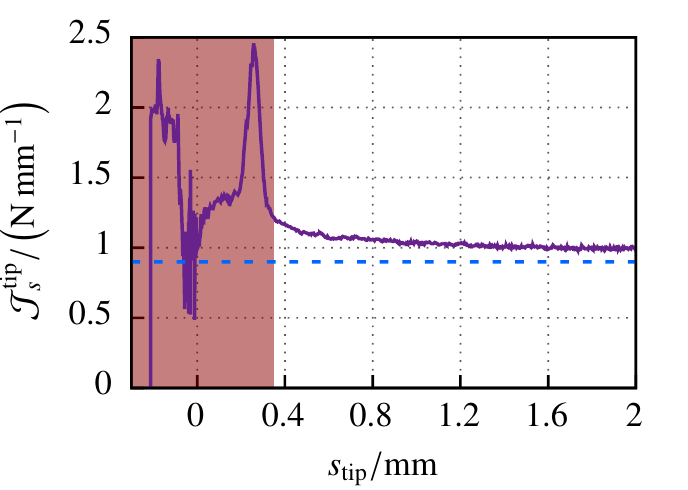}
		\label{fig:quantAnIncl_err}
	}
	\subfloat[Crack tip trajectory, $\gcb/\gci=3$]{
		\includegraphics{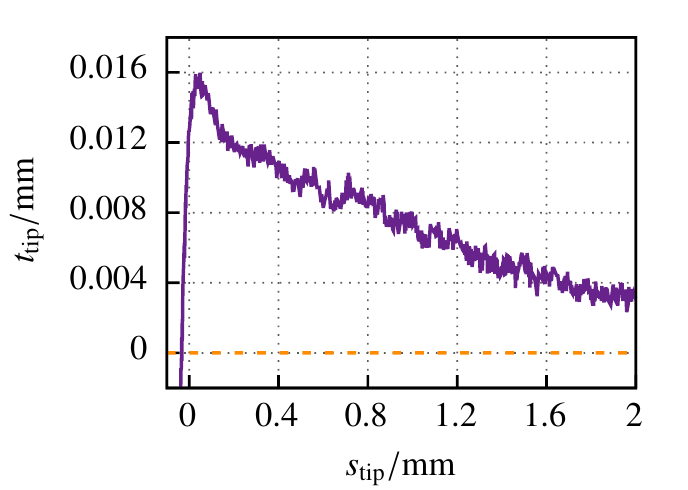}
		\label{fig:quantAnIncl_tip}
	}
	\caption{Energy release rate \textbf{(a)} and crack tip trajectory \textbf{(b)} for a crack which deflects into the interface. $\jt_i$ is transformed from $i=x,y$ to $i=s,t$, where the $s$- and $t$-directions are aligned with and perpendicular to the interface, respectively. The value $\jt_s$ is quantitatively compared to the interface fracture toughness \legline{6}. 
	Within the red interval, $\jt_s$ determined from the configurational force balance does not recover the actual crack driving force. Outside the interval, there is good agreement between the energy release rate at which the crack propagates and the fracture toughness.
	The corresponding crack tip trajectory in terms of the $s,t$-system is shown in \textbf{(b)}, where \legline{8} marks the interface midline. The crack is, as expected, deflected into the interface along the $s$-direction. }
	\label{fig:quantAnIncl}
\end{figure}

The numerical predictions for the inclined interface are compared quantitatively considering the energy release rate in the direction of crack growth and the crack tip trajectory. Both are depicted in Figure~\ref{fig:quantAnIncl} for a crack which propagates along the interface midline for $\phii=\SI{45}{\degree}$ and $\gcb/\gci=3$. Therefore, a transformed $s,t$-coordinate frame, aligned with the interface midline, is introduced, see Figure~\ref{fig:contIncl_defl3}.

As mentioned above, the crack tip follows a curved path when it deflects into the interface. 
From the tip coordinates $[s_\text{tip}\,\, t_\text{tip}]\T$ transformed into the coordinate frame aligned with the interface, it becomes clear, that the crack tip trajectory looks quite similar to the path which is observed for a crack deflected into a perpendicular interface, compare Figure~\ref{fig:quantAnIncl_tip} to Figure~\ref{fig:quantAn_br_tip}.
When the crack tip deflects, it first overshoots the interface midline, but approaches the midline when it continues to propagate along the interface. 
For the inclined interface, the distance between interface midline and actual position of the crack tip which propagates along the interface and hence the uncertainty of the crack tip position that stems from the regularization and the crack tip tracking method is slightly higher than for the perpendicular interface.
Nevertheless, it does not exceed $\elli$, which is again deemed acceptable in the context of the regularized framework. 

Figure~\ref{fig:quantAnIncl_err} shows the corresponding energy release rate in direction of the interface midline $\jt_s$.
Similar to the perpendicular interface, $\jt_s$ which is determined from the balance of the configurational forces does not correspond to the crack driving force at every instant.
Instead, for $s_\nv{tip} < r = \SI{0.35}{\milli\meter}$, the validity of $\jt_s$ is compromised, because the area in which the crack deflects into the interface is located within the integration domain $\mathcal{C}$ of the configurational forces.
The corresponding interval is marked in red in Figure~\ref{fig:quantAnIncl_err}.
Only for $s_\nv{tip} > r = \SI{0.35}{\milli\meter}$, $\jt_s$ recovers the actual crack driving force.
Similar to the perpendicular case, the energy release rate of the crack propagating along the interface approximately meets the interface fracture toughness $\gci$, yet is slightly higher than the exact value.
As it has been outlined in Section~\ref{sec:elhomperp}, this slight overestimation is mainly caused by the uncertainty of the position of the crack tip which arises from the regularization.

\subsection{Heterogeneous elasticity and inclined interface}

\begin{figure}[t]
	\centering
	\subfloat[$\gcb/\gci=1.3$]{
		\includegraphics{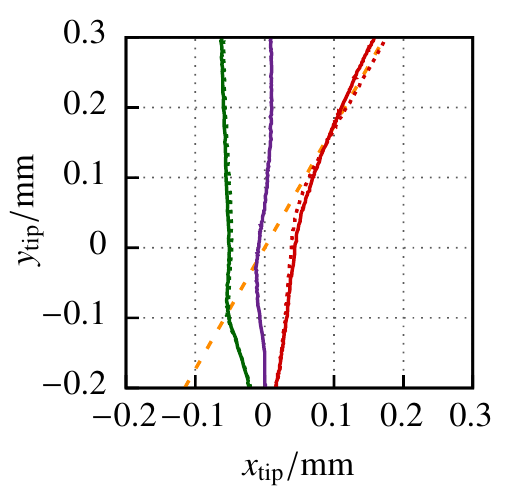}
		\label{fig:contInclHet_defl1}
	}
	\subfloat[$\gcb/\gci=1.8$]{
		\includegraphics{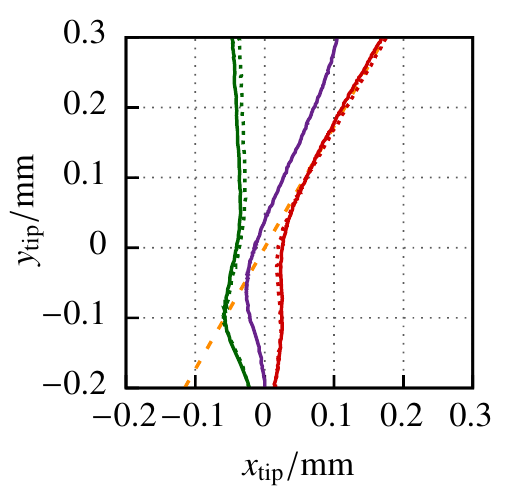}
		\label{fig:contInclHet_defl2}
	}
	\subfloat[$\gcb/\gci=2.2$]{
		\includegraphics{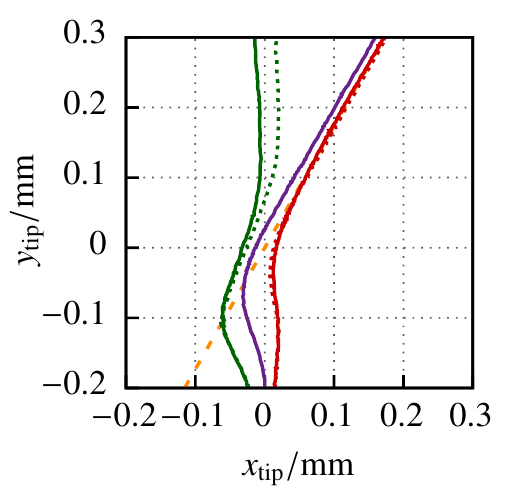}
		\label{fig:contInclHet_defl3}
	}
	\\
	\vspace{0.3cm}
	\footnotesize 
	\begin{tabular}{l|c|c|c}
	&$\alpha=-0.5$  & $\alpha=0$ & $\alpha=0.5$ \\
	Trajectories for $\rtan$ &	\legline{1}  & \multirow{2}{*}{\legline{17}}  & \legline{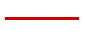} \\   
	Trajectories for $\reheavi$ &	\legline{2} &  & \legline{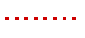} \\  
	\end{tabular}
	\caption{Nine different crack phenomena, three for each ratio $\gcb/\gci$ \textbf{(a)}~--~\textbf{(c)}, for an initial crack which propagates towards an inclined interface $\phii=\SI{30}{\degree}$, depicted by the midline \legline{8}. The \textsc{Young}'s modulus varies across the interface according to $\alpha$ and the crack tip trajectories differ accordingly.
	Firstly, comparing the trajectories for $\alpha=0$ to the ones from Figure~\ref{fig:contIncl}, it is evident that a smaller inclination angle $\phii$ yields concurrent phenomena for lower ratios $\gcb/\gci$ which is intuitive.
	Secondly, the elastic heterogeneity strongly influences the results according to the LEFM predictions~\cite{he_crack_1989}, which stated that for $\alpha>0$, the crack has a tendency away from the interface and vice versa. Thirdly, total deflection into the interface can be preferred or delayed depending on $\alpha$.
	}
	\label{fig:contInclHet}
\end{figure}

In a final numerical study, the impact of an elastic heterogeneity on the fracture phenomena at an inclined interface is investigated. The computed crack tip trajectories for $\phii=\SI{30}{\degree}$, $\alpha \in \{-0.5, 0, 0.5\}$ and $\gcb/\gci \in \lbrace 1.3, 1.8, 2.2 \rbrace$ are depicted in Figure~\ref{fig:contInclHet}. They serve as representative examples for all phenomena which were predicted numerically, because the same qualitative influence of $\alpha$ was observed for different interface inclination angles $\phii$ and fracture toughness ratios $\gcb/\gci$.
The elastic heterogeneity was described using $\rtan$. Simulations were also carried out for a mesh-conforming jump $\reheavi$ of the \textsc{Young}'s modulus with respect to the interface midline for the parameter set mentioned above.
All other parameters are as in the previous sections.

Firstly, the trajectories for $\alpha=0$ in Figure~\ref{fig:contInclHet} are compared to Figure~\ref{fig:contIncl}, which presents the same three types of crack phenomena for a different angle $\phii=\SI{45}{\degree}$. As expected, the three phenomena, a straight crack (a), a small deflection (b) and distinct interface failure (c), occur for lower ratios $\gcb/\gci$ for a smaller angle $\phii=\SI{30}{\degree}$.

Secondly, the elastic heterogeneity has an influence on the crack path when a crack approaches the interface. For $\alpha < 0$, the crack even more deflects in the direction of the interface than for $\alpha = 0$. On the contrary, it tends away from the interface for $\alpha > 0$. In other words, the crack tends to propagate away from the interface when the material beyond the interface is stiffer than the one in front of the interface and follows a curved path towards the interface, otherwise.
In additional simulations which are not reported here, it has been observed that this effect becomes the more pronounced for a larger elastic dissimilarity between the two bulk materials. The deviation towards the interface for $\alpha < 0$ and vice versa is consistent with the LEFM~\cite{he_crack_1989}. For a \textit{wedge-loaded} crack approaching an inclined interface from a finite distance, a curved path in the direction of the interface and away from the interface has been predicted analytically for $\alpha < 0$ and $\alpha >0$, respectively.

Thirdly, the elastic heterogeneity controls whether deflection into the interface occurs. The corresponding critical fracture toughness ratio decreases with increasing $\alpha$ and the crack length along the interface becomes higher when $\alpha$ increases, respectively.
This is consistent with the simulations and LEFM predictions \cite{he_crack_1989} for the perpendicular interface, see Section~\ref{sec:hetel-per}.

Similar to the investigations in Section~\ref{sec:hetel-per}, the results for a smooth transition of the \textsc{Young}'s modulus $\rtan$ are compared to a mesh-conforming jump $\reheavi$. The according crack tip trajectories are depicted in Figure~\ref{fig:contInclHet} as dotted lines.
Although the results qualitatively agree, deviations exist which have to be quantified in further studies, where a \textit{partial rank-I relaxation} is implemented. 

\section{Conclusion}
\label{sec:conlc}

A phase-field model for brittle fracture has been presented which incorporates materials with dissimilar elastic properties and interfaces between them in a regularized manner.
The discrete interface is regularized over a finite length by means of the finite interface regularization length scale $\elli$.
Since this length scale is very small compared to the domain's dimension, the interface is called \textit{narrow}. It was observed in previous studies, that the characteristic length of the crack phase-field model $\ellc$ and the length scale $\elli$ exhibit an interaction, i.e. the phase-field profile of an interfacial crack is altered with respect to the case of a homogeneous fracture toughness.
Furthermore, the dissipated energy related to a crack along the interface is overestimated due to an influence of the bulk material. A numerical compensation approach was adopted that overcomes this issue and yields crack propagation along the interface energetically independent from the crack and interface regularization length scales, and from the fracture toughness of the bulk material.
The compensation was analytically motivated considering the one-dimensional case. 

Materials adjacent to the interface may have dissimilar elastic properties. The model introduces a smooth transition of the sharp variation by a hyperbolic tangent function, which alters the elastic properties accordingly and is controlled by the interface length scale $\elli$, too. 

The modeling framework was validated against analytical analyses from He and Hutchinson~\cite{he_crack_1989}. They investigated crack branching and deflection phenomena for a crack, which impinges a possibly inclined interface. For the first of the four investigated setups, three different interface regularizations were investigated. Due to numerical issues and the general ability to predict all crack patterns, the \textsc{Gaussian}-like fracture toughness regularization was chosen for the subsequent investigations. In the course of the regularization function comparison, the interface length scale $\elli$ proved to be a parameter which is not of numerical nature but rather a material parameter. Additionally, the phenomenon of a single deflection as described by He and Hutchinson~\cite{he_crack_1989} was never observed. This may be due to the fact that the tensile split reveals significant disadvantages when it comes to shear load cases~\cite{steinke_phase-field_2018}. Despite its disadvantages and limitations, it is widely used in the phase-field community because of its intriguing simplicity compared to physically-based approaches.
However, for a correct prediction of cracking phenomena under manifold loading conditions, it is necessary to switch to a different directional split as for example introduced by Steinke and Kaliske~\cite{steinke_phase-field_2018}, who degrade the stress according to a local crack coordinate frame. 

Next, the three remaining setups on the basis of analytical investigations from He and Hutchinson~\cite{he_crack_1989} were considered. The qualitative agreement between the LEFM predictions and the present numerical investigations like the dependence of the crack pattern on the ratio $\gcb/\gci$, on \textsc{Dundurs}' parameter $\alpha$ and the interface inclination angle $\phii$ are good. Numerically, a smooth transition between different phenomena was observed, for instance for the inclined interface, whereas the analytical predictions are of binary nature. A quantitative evaluation of the crack tip trajectory and the configurational forces served as evidence for the crack driving forces of different crack patterns. Comparing the analytically predicted switching ratios between different phenomena to the corresponding numerical simulations is not straightforward because of the smooth transition between different phenomena, especially for the inclined interface. A quantification is therefore impossible but the order of magnitude of the relevant ratios corresponds in principle. Comparative simulations with a sharp jump of the \textsc{Young}'s modulus revealed, that the hyperbolic tangent function which describes the elastic heterogeneity significantly influences the results. The chosen interpolation scheme between the \textsc{Young}'s moduli does not necessarily fulfil the mechanical jump conditions as outlined in~\cite{schneider_phase-field_2015,kiefer_numerical_2017}. A remedy to this issue is for example the \textit{partial rank-I relaxation} as discussed in~\cite{mosler_novel_2014,schneider_phase-field_2015,kiefer_numerical_2017}, which is going to be extended to be applied in the model above in the future. Despite a quantitative disagreement for certain simulation results, the presented model already captures most effects and serves as sound basis for further development and investigations. 


\section*{Data Availability}
The raw data for the compensation curves in Figure~\ref{fig:compensation} is provided upon request. Please cite this article, if reused in any form.

\section*{Acknowledgements}
The authors gratefully acknowledge support by the Deutsche Forschungsgemeinschaft in the Priority Program 1748 “Reliable simulation techniques in solid mechanics. Development of non-standard discretization methods, mechanical and mathematical analysis” under the project KA3309/3-2, and from the European Research Council under Advanced Grant PoroFrac (grant number 664734). The computations were performed on a HPC-Cluster at the Center for Information Services and High Performance Computing (ZIH) at TU Dresden. The authors thank the ZIH for allocations of computational time. Finally, the authors thank all three referees for the constructive comments.


\appendix

\section{Implementation in \fenics}
\label{sec:impl}
As supplementary material, we provide a Python code example which can be used as reference for the implementation of the presented model. It can be found in the online version, at \url{https://doi.org/10.1016/j.engfracmech.2020.107004} 

For the implementation of the tensile split, a UFL-compliant formulation for the spectral decomposition of second order tensors is presented because the \textit{Unified Form Language} does not provide any functionality regarding eigenvalues and -vectors by default.

In the following, the key aspects for the formulation of the spectral decomposition are outlined.
For the eigenvalues $e_i$ of the strain tensor, the analytical solution of the characteristic polynomial
\begin{equation}
  e_i^3 - I_1 \, e_i^2 + I_2 \, e_i - I_3 = 0
\end{equation}
is implemented, where
\begin{equation}
 I_1 = \tr \te{\varepsilon} \coma \hspace{2mm}
 I_2 = \frac{1}{2}\left[ \tr^2 \te{\varepsilon} - \tr \left(\te{\varepsilon} \cdot \te{\varepsilon} \right) \right] \coma \hspace{2mm}
 I_3 = \det \, \te{\varepsilon}
\end{equation}
designate the principal invariants of $\te{\varepsilon}$. 
For this purpose, \textsc{Cardano}'s rule is exploited, cf. Kopp \cite[Equations (19)~--~(34)]{kopp_efficient_2008}.
Numerical difficulties arise if the weak form implemented in UFL leads to the numerical evaluation of terms exhibiting singular or complex values.
Consequently, special attention has to be paid to fractions and square roots which are evaluated while the automatically generated code is executed.
In order to avoid computational complications, small \textit{numerical perturbations} are added to all radicands and denominators showing an absolute value lower than a critical threshold, see \cex{lst:specdec},~ll.~8~--~10. As both the \textit{numerical perturbations} and the threshold are set to values in the order of $10^{-10}$, this strategy is assumed to have no remarkable impact on the simulation results.

\begin{lstlisting}[float=h,caption={Concept of \textit{numerical perturbations} for the implementation of the spectral decomposition},label={lst:specdec}]
nuz = 1E-10 # threshold for application of numerical perturbations
numturb = nuz # numerical perturbations applied e.g. in case of equal eigenvalues

# ...
def eval1(te): # first eigenvalue of tensor te	
# ...

inroot1=conditional(inroot1<nuz,abs(inroot1)+numturb,inroot1)

denom1=conditional(abs(denom1)<nuz,conditional(denom1<0,denom1-numturb,denom1+numturb),denom1)

#...    
# ...
def M1(te): # projection tensor associated to first eigenvalue
eva1=eval1(te)
# eva2,	eva3 = ... other eigenvalues

# ensure numerically different eigenvalues
eva1=conditional(abs(eva1-eva2)<nuz,eva1+numturb,eva1) 
eva1=conditional(abs(eva1-eva3)<nuz,eva1+2*numturb,eva1)
eva2=conditional(abs(eva2-eva3)<nuz,eva2+4*numturb,eva2)

denom1=(eva1-eva2)*(eva1-eva3)

return (1.0/denom1)*dot((te-eva2*Identity(sd)),(te-eva3*Identity(sd)))
\end{lstlisting}

With the eigenvalues $e_i$ at hand, the spectral decomposition
\begin{equation}
\te{\varepsilon} = \sum_{i \in [1,3]} e_i \hspace{2mm} \te{M}_i
\label{eq:specdc-projtens}
\end{equation}
can be formulated making use of the projection tensors
\begin{equation}
\te{M_i} =\frac{1}{D_i} \, \prod_{j \in [1,3] \setminus i} \left( \te{\varepsilon} - e_j \, \te{1} \right)
\hspace{5mm} \text{with} \hspace{5mm}
 D_i = \prod_{j \in [1,3] \setminus i} \left( e_i - e_j\right) \coma
\label{eq:projtens}
\end{equation}
which are also referred to as \textit{eigenvalue basis} in the literature, cf. Miehe~\cite{miehe_comparison_1998}.
Since Equation~\eqref{eq:projtens} is only applicable in cases of three pairwise different eigenvalues, the concept of \textit{numerical perturbations} is exploited again as demonstrated in \cex{lst:specdec},~ll.~19~--~21.


\section*{References}
\bibliographystyle{elsarticle-num} 
\bibliography{literatur.bib}





\end{document}